\documentclass[english, prl, showpacs, onecolumn,
reprint,
superscriptaddress
]{revtex4-2}
\usepackage[dvipsnames]{xcolor}
\usepackage{babel}
\usepackage{verbatim}
\usepackage{bm}
\usepackage{amsmath}
\usepackage{amssymb}
\usepackage{graphicx}
\usepackage{tabularx}
\usepackage{multirow}
\usepackage{booktabs}
\usepackage{colortbl}
\usepackage{float, extpfeil}
\usepackage{placeins}
\usepackage{makecell}

\definecolor{tabcolor}{rgb}{.410,.10,.11}
\usepackage[unicode=true, pdfusetitle,
 bookmarks=true, bookmarksnumbered=false, bookmarksopen=false,
 breaklinks=false, pdfborder={0 0 0}, pdfborderstyle={}, backref=false, colorlinks=true]
 {hyperref}
\hypersetup{citecolor=blue}

\usepackage{soul}
\usepackage{ulem}

\soulregister\cite7
\soulregister\ref7
\soulregister\pageref7

\makeatletter

\begin{document}

\author{Jia-Nan Wu}
\affiliation{Beijing Key Laboratory of Nanophotonics $\&$ Ultrafine Optoelectronic Systems, School of Physics, Beijing Institute of Technology, Beijing 100081, China.}
\author{Bingsuo Zou}
\affiliation{MOE $\&$ Guangxi Key Laboratory of Processing for Non-ferrous Metals and Featured Materials, School of Physical science and Technology, Guangxi University, Nanning 530004, China.}
\author{Yongyou Zhang}
\email[Electronic mail: ]{yyzhang@bit.edu.cn}
\affiliation{Beijing Key Laboratory of Nanophotonics $\&$ Ultrafine Optoelectronic Systems, School of Physics, Beijing Institute of Technology, Beijing 100081, China.}

\title{Statistics of tens-of-photon states scattered by optical cavity, two-level atom and Jaynes-Cummings emitter}

\date{\today} 

\begin{abstract}
Manipulating photon states serves as a primary requirement for various optical devices and is of high relevance for quantum information technology. Nevertheless, the fundamental theoretical framework for tens-of-photon states has not been established. This study successfully establishes the matrix-product-state theory to explore the statistics of the tens-of-photon states scattered by optical cavities (OCs), two-level atoms (TLAs), and Jaynes-Cummings emitters (JCEs) in waveguide-QED systems. Taking 10-photon states as an example, we reveal some novel physical results that differ from those for few-photon cases. We verify that OCs do not change the statistics of the incident photon states, being independent of the photon number. However, for the TLAs and JCEs, the photon number strongly impacts the photon bunching and anti-bunching behaviors. As the photon number increases, there exists a maximum strength for the photon-photon correlation induced by the JCE. Especially, the scattered waves by the TLA (or JCE) exhibit extremely different statistics behaviors for the 10-photon cases from those for the bi-photon. These distinguishable conclusions for the tens-of-photon states and the matrix-product-state theory pave the way for the multi-photon manipulation.
\end{abstract}

\maketitle

{\sl Introduction.---}The dynamics of quantum many-body systems \cite{vidal2004efficient, eisert2015quantum, carleo2017solving, weimer2021simulation, felser2021efficient} present intriguing challenges that consistently captivate researchers' attentions.
Notably, the intricate dynamics of multi-photon phenomena are extensively explored \cite{shen2007strongly, yudson2008multiphoton, pan2012multiphoton, baragiola2012n, shi2015multiphoton, Kocaba2016Effects, liao2020multiphoton} due to photons being optimal carriers for quantum information.
Manipulating photons stands as a pivotal technology for optical quantum computers \cite{zhong2020quantum}.
The realm of waveguide-QED systems \cite{shen2009theory, Liao_2016, sheremet2023waveguide}, which encompasses one-dimensional (1D) waveguides and quantum emitters like optical cavities (OCs) \cite{waks2006dipole, shi2011two, li2023single}, two-level atoms (TLAs) \cite{rephaeli2011few}, and Jaynes-Cummings emitter (JCEs)\cite{shen2009theory} containing an OC and an inside TLA, provides a straightforward yet effective platform for the study of the multi-photon manipulation.

Building upon these systems, researchers have contributed a lot of theoretical \cite{shen2007strongly, yudson2008multiphoton, shi2011two, rephaeli2011few, pan2012multiphoton, baragiola2012n, shi2015multiphoton, Kocaba2016Effects, Hu2018Transmission, liao2020multiphoton}
and experimental \cite{2008Generation, matthews2009manipulation}
works, pertaining to multi-photon transport \cite{2008Generation, rephaeli2011few, shi2011two, Liao_2016, Hu2018Transmission}, correlations \cite{shen2015photonic, Kocaba2016Effects, Hu2018Transmission}, bunching and anti-bunching effects \cite{shen2015photonic}, entanglement \cite{matthews2009manipulation, Miraz2016Two}, bound states \cite{zheng2010waveguide}, and more.
In these reported works, researchers have proposed the Bethe-ansatz approach \cite{shen2007strongly}, Lehmann-Symanzik-Zimmermann reduction \cite{PhysRevB.79.205111, shi2011two, Shi2013Two}, Green's function decomposition of multiple particle scattering matrices \cite{see2017diagrammatic}, input-output formalism \cite{Fan2010Input, 6189728, Xu2015Input, PhysRevA.101.063809}, Feynman diagram formalism \cite{shi2011two}, and so on.
According to these methods, the statistics for the few-photon states scattered by quantum emitters like OCs \cite{shi2011two}, TLAs \cite{shi2015multiphoton}, and JCEs \cite{shi2015multiphoton, 6189728} have been studied. As for the few-photon cases, the TLA results in a bunching in the transmitted state while in an anti-bunching in the reflected state \cite{shen2015photonic}. 
The OCs do not change the statistics behaviors of the incident photon states \cite{shi2011two}.

Recalled the philosophy of ``more is different'' \cite{anderson1972more} by P. W. Anderson, novel phenomena should appear in the scattering process of the states with much more than one photons.
Regretfully, the methods listed above only suit the systems involving no more than four photons. The main challenge they face is the exponential growth of the Hilbert spaces. To overcome it, we turn to the matrix product state (MPS) theory \cite{SCHOLLWOCK201196, PAECKEL2019167998, RevModPhys.93.045003}, which draws inspiration from the principles of the density-matrix renormalization group \cite{White1992Density, PhysRevB.48.10345} and offers an intelligible framework for studying 1D quantum many-body states \cite{SCHOLLWOCK201196, PAECKEL2019167998}.
The MPS theory has been adeptly employed in examining the temporal evolutions of strong-correlation systems \cite{Johnson2010Dynamical, Singh2014Matrix, PAECKEL2019167998}, particularly in the context of 1D spin chains \cite{vidal2004efficient, Mazza2019Suppression, Friedman2022Topologicaledge} and the Bose-Hubbard model \cite{URBANEK2016170}.

To solve the statistics of tens-of-photon states in waveguide-QED systems, this letter firstly establishes the MPS theory and then takes 10-photon states as examples to explore the multi-photon statistics, scattered by the OCs, TLAs, and JCEs.
According to them, we verify that the OCs do not change the statistics of the incident photon states, being independent of the photon number.
However, the photon number strongly impacts the photon bunching and anti-bunching behaviors for the transmitted and reflected pulses scattered by the TLAs and JCEs. 
In addition, we demonstrate that there exists a maximum strength for the photon-photon correlation induced by the JCEs from the perspective of transmission. 
In particular, the scattered states by the TLA (or JCE) exhibit extremely different statistics behaviors between the bi-photon and 10-photon cases, including the second-order correlation.

\begin{figure}
  \centering
  \includegraphics[width=0.44\textwidth]{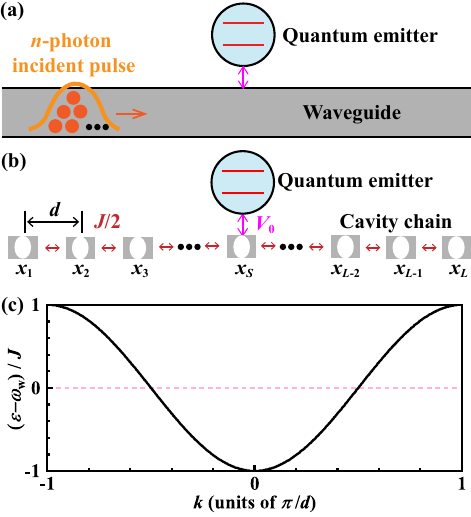}\\
  \caption{(a) Schematic of a quantum emitter side-coupled to an infinite waveguide. The incident multi-photon pulse is depicted by the orange curve and solid dots. (b) Discretized model of (a), with $L$ cavities. The position of the $l$th cavity is $x_l=(l-1)d$, with lattice constant $d$. $J/2$ and $V_0$ denote the hopping between adjacent cavities and the coupling between the quantum emitter and the $S$th cavity, respectively. (c) Dispersion of the cavity chain in (b)\cite{PhysRevA.96.023831, Qiao2019Quantumphase}.}\label{Fig1}
\end{figure}

{\sl Theoretical framework.---}We use the waveguide-QED system drawn in Fig.~\ref{Fig1}(a) as cornerstone to present the way of  establishing the MPS framework. 
Figure~\ref{Fig1}(a) depicts a 1D waveguide side-coupled to a quantum emitter, where the incident $n$-photon pulse is denoted by the orange curve and solid dots.
To meet the requirements of the MPS framework, the waveguide is discretized into a cavity chain, see Fig.~\ref{Fig1}(b), with the site number $L$, lattice constant $d$, and nearest neighbor hopping $J/2$.
All the cavities hold the identical eigenfrequencies $\omega_W$.
The $S$th cavity is coupled to the quantum emitter. The chain dispersion is $\varepsilon=\omega_W-J\cos(kd)$ \cite{PhysRevA.96.023831, Qiao2019Quantumphase}, see Fig.~\ref{Fig1}(c).
Such a lattice is described by the Hamiltonian,
\begin{align}
\hat{H}=\hat{H}_{W}+\hat{H}_{E}+\hat{H}_{I},
\end{align}
where $\hat{H}_{W}$, $\hat{H}_{E}$ and $\hat{H}_{I}$, respectively, correspond to the cavity chain, quantum emitter and interaction between them. They read
\begin{align}\label{HW}
\hat{H}_{W}&=\sum_{l=1}^{L}\omega_{W}\hat{a}_{l}^{\dag}\hat{a}_{l}-\frac{J}{2}\sum_{l=1}^{L-1}\left(\hat{a}_{l}^{\dag}\hat{a}_{l+1}+\rm h.c.\right),\\
\hat{H}_{E}&=\omega_c \hat a_c^\dag \hat a_c + \omega_{a}\hat{\sigma}^{+}\hat{\sigma}^{-} + \Omega\left(\hat a_c^\dag\sigma^{-} + \sigma^{+}\hat a_c\right),\\
\hat{H}_{I}&=V_{0}\left(\hat{a}_{S}^{\dag}\hat{a}_{c}+\hat{a}_{c}^{\dag}\hat{a}_{S}\right).
\end{align}
Here, the Planck constant is set to $\hbar=1$ for convenience.
$\hat{a}_{l}^{\dag}$ ($\hat{a}_{l}$) represents the creation (annihilation) operator of the $l$th cavity in the chain.
$\hat{H}_{E}$ takes the JCE as an example, where $\hat{a}_c^{\dag}$ ($\hat{a}_c$) is the creation (annihilation) operator of the cavity in the JCE and $\sigma^{+}$ ($\sigma^{-}$) is the raising (lowering) operator of the corresponding TLA. $\Omega$ describes the Rabi coupling between them.
$V_0$ measures the coupling between the cavity in the JCE and the $S$th cavity in the chain.

To apply the MPS method to this discretized waveguide-QED model, we write the multi-photon states into the MPS form,
\begin{align}\label{mps}
\left|\Phi\right\rangle_{n}=
\sum_{\{\bm \sigma\}}M^{[n]\sigma_{1}}_{1}M^{[n]\sigma_{2}}_{2}\cdots M^{[n]\sigma_{L}}_{L}\left|\sigma_{1}\sigma_{2}\cdots\sigma_{L}\right\rangle,
\end{align}
where $\sigma_l$ is the occupation number on the $l$th site and $n$ is the total one with $\sigma_l\leq n$.
$\{\bm \sigma\}$ represents the set of $\{\sigma_1,\sigma_2,\cdots, \sigma_L\}$.
The matrix of $M^{[n]\sigma_l}_{l}$ rests on the $l$th site, see Fig.~\ref{Fig1}(b), where $[n]$ means that there are $n$ excitations in $\left|\Phi\right\rangle_{n}$.
$M^{[n]\sigma_{1}}_{1}$ is a one-row matrix, while $M^{[n]\sigma_{L}}_{L}$ is a one-column one.
We take the following $n$-photon Gaussian pulse,
 \begin{align}\label{ini_st}
\left|i\right\rangle_{n}= \sum_{l_1,l_2,\cdots, l_n}
\phi_{l_1}\phi_{l_2}\cdots \phi_{l_n}
\hat a_{l_1}^\dag \hat a_{l_2}^\dag \cdots \hat a_{l_n}^\dag
\left|\varnothing\right\rangle,
\end{align}
as an example, where $\left|\varnothing\right\rangle$ is the vacuum state.
The coefficient $\phi_{l}$ reads
$
\phi_l=N\sum_{k}e^{-(k-k_{0})^{2}/k_{w}^{2}}e^{-ik(x_l-x_0)},
$ with the normalized constant $N$ and the wave vector $k={2\pi\over L}l$ ($l=1, 2, \cdots, L$).
This $\phi_l$, with the central wave vector $k_{0}$, the central energy $\varepsilon_0=\omega_W-J\cos{k_0d}$, the central position $x_{0}$, and the width $2k_w^{-1}$, guarantees the invariance of $\left|i\right\rangle_{n}$ under the exchange of any two photons. Note that $\left|i\right\rangle_{n}$ is a Fock state.

It is not a trivial task to transform $|i\rangle_n$ into the form of Eq.~\eqref{mps}, especially for the cases with large $n$ and $L$, because of the exponentially increase of the Hilbert space.
The relation $\left|i\right\rangle_{n} = \left(\sum_{l_n}{\phi_{l_n}}\hat a_{l_n}^\dag\right)\left|i\right\rangle_{n-1}$ suggests us to establish an iterative relation between the MPS forms of $\left|i\right\rangle_{n}$ and $\left|i\right\rangle_{n-1}$.
The procedure \cite{SI} includes 
(i) writing the single-occupation state $\phi_{l_1}\hat a_{l_1}^\dag|\varnothing\rangle$ into the MPS form, 
(ii) constructing the MPS form of $|i\rangle_1$ by superposing the MPSs of all $\phi_{l_1}\hat a_{l_1}^\dag|\varnothing\rangle$, 
(iii) deriving the iteration relation between the MPSs of ${\phi_{l_n}}\hat a_{l_n}^\dag\left|i\right\rangle_{n-1}$ and $\left|i\right\rangle_{n-1}$, 
and (iv) similar to (ii), constructing the MPS form of $|i\rangle_n$ by superposing the MPSs of all ${\phi_{l_n}}\hat a_{l_n}^\dag\left|i\right\rangle_{n-1}$. 
Note that the physical freedom of the quantum emitter can be neglected temporally in the procedure of finding the MPS of $\left|i\right\rangle_{n}$, because it can be added by expanding the physical dimension of the $S$th site \cite{SI}.

We next construct the MPO for the time-evolution operator $e^{-i\hat H\tau}$ with a short time step $\tau$, i.e.,
\begin{align}
\hat{O}_\tau^{[n]}=\sum_{\{\bm \sigma, \bm \sigma'\}}
Q_{1}^{[n]\sigma_{1}\sigma'_{1}}Q_{2}^{[n]\sigma_{2}\sigma'_{2}}
\cdots
Q_{L}^{[n]\sigma_{L}\sigma'_{L}} \qquad \nonumber\\
\times
\left|\sigma_{1}\sigma_{2}\cdots\sigma_{L}\right\rangle
\left\langle\sigma'_{1}\sigma'_{2}\cdots\sigma'_{L}\right|,
\end{align}
where $Q_{l}^{[n]\sigma_{l}\sigma'_{l}}$ is a matrix with $0\leq\sigma_l, \sigma_l' \leq n$.
The second-order Trotter decomposition is used, that is,
$
e^{-i\hat H\tau} \approx
e^{-i\hat H_{\rm odd}\tau/2}
e^{-i\hat H_{\rm even}\tau}
e^{-i\hat H_{\rm odd}\tau/2}
$ \cite{SCHOLLWOCK201196},
where $\hat H_{\rm odd}$ and $\hat H_{\rm even}$ are the Hamiltonians on the odd and even bonds, respectively.
Let's denote the Hamiltonian on the $l$th bond as $\hat{h}_{l}$.
Since $[\hat{h}_{l},\hat{h}_{l+2}]=0$, the equations
$
e^{-i\hat{H}_{\rm odd}\tau/2}=e^{-i\hat{h}_{1}\tau/2}e^{-i\hat{h}_{3}\tau/2}\cdots
$
and
$
e^{-i\hat{H}_{\rm even}\tau}=e^{-i\hat{h}_{2}\tau}e^{-i\hat{h}_{4}\tau}\cdots
$
are strict, whose last terms is determined by the $L$'s parity.
The way for constructing the MPO of $e^{-i\hat{h}_{l}\tau}$ is shown in \cite{SI}.

The time evolution of the MPS can be implemented by iteratively applying the MPO of $e^{-i\hat H\tau}$ on $|i\rangle_n$ \cite{SI}.
The calculation errors primarily stem from Trotter decomposition and the truncation of both the MPS and MPO \cite{suzuki1976generalized, Singh2014Matrix, PAECKEL2019167998}.
They can be mitigated by decreasing the time step and increasing the matrix dimensions of the MPS and MPO.
The accuracy in the following calculations is assessed by monitoring the deviations of the total occupation numbers from their initial values, which always remain below $0.01$.

\begin{figure}
  \centering
  \includegraphics[width=0.48\textwidth]{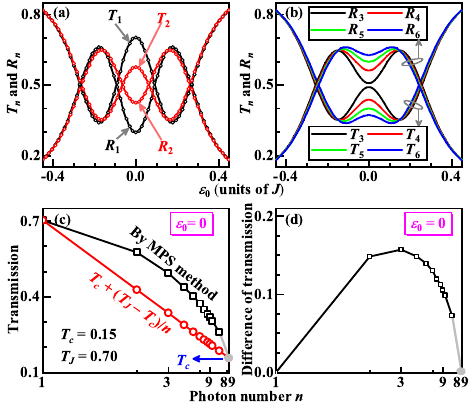}\\
  \caption{Transmission ($T_n$) and reflection ($R_n$) for $n$-photon Gaussian pulses.
(a) Lines are calculated by the conventional Schr\"odinger equation, on which the circular dots are by the MPS method.
(b) Spectra calculated by the MPS method for $n=3, 4, 5,\rm and\ 6$.
(c) Variation of the transmission at $\varepsilon_0=0$ with $n$ (square black dots), with respect to that of $T_c+(T_J-T_c)/n$ (circle red dots).
(d) Difference between the square black and circle red dots in (c). The grey lines and dots in (c) and (d) show the limitation of $n\rightarrow\infty$.
Parameters: $L=256$, $S=128$, $k_{w}=0.05\pi/d$, $x_{0}=64d$, $\omega_{a}=\omega_c=\omega_{W}\equiv0$, $V_0=0.4J$, and $\Omega=0.15J$.}\label{Fig2}
\end{figure} 

{\sl Transmission spectra.---}Before applying the MPS theory to tens-of-photon processes, we verify its effectiveness in the few-photon cases, that is, single- and bi-photon pulses scattered by a JCE, see Fig.~\ref{Fig2}(a).
The solid lines are obtained using the Schr\"odinger equation \cite{shen2007strongly, Hu2018Transmission}, $i\hbar{\partial \over\partial t}|\psi(t)\rangle=\hat H |\psi(t)\rangle$.
The circular dots represent the corresponding values obtained by the MPS theory.
The two methods yield consistent results.

\begin{figure*}
  \centering
  \includegraphics[width=1\textwidth]{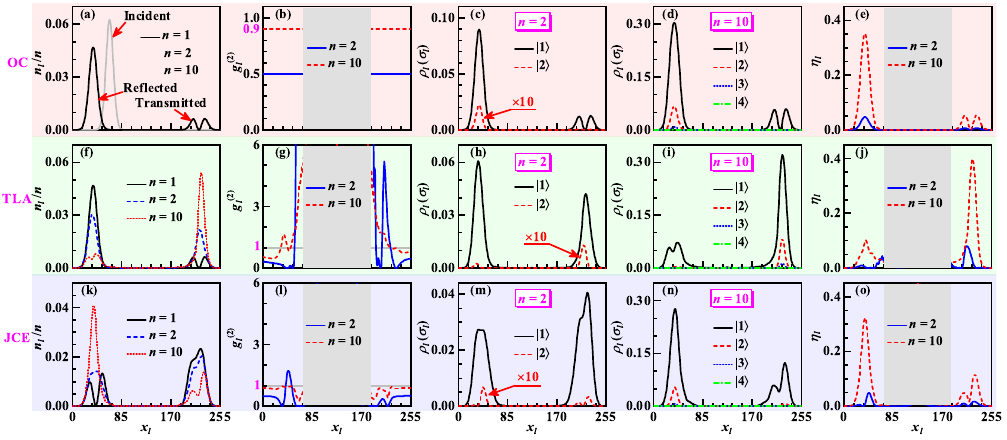}\\
  \caption{Statistics of the final states.  (a, f, k) Distributions of normalized photon numbers for $n=1$, 2, and 10.
  (b, g, l) Normalized second-order correlation functions for $n=2$ and 10.
 Occupation probabilities of the Fork states $|\sigma_l\rangle$ for $n=2$ (c, h, m) and $n=10$ (d, i, n).
  (e, j, o)  Probabilities of the photons coming from the multi-occupation states.
  The red dashed lines in (c, h, m) are magnified 10 times.
  The quantum emitter used in each row is denoted on the left side.
  The left and right regions in each panel correspond to the reflected and transmitted pulses, see (a).
  Parameters are same with those in Fig.~\ref{Fig2} with $\varepsilon_0=0$.}\label{Fig3}
\end{figure*}

Since it has become a challenge in calculation even as $n=3$ (whose Hilbert space dimension reaches 2 796 416) using the conventional methods \cite{shen2007strongly, shi2011two, PhysRevB.79.205111, see2017diagrammatic, 6189728}, the transmission and reflection spectra in Fig.~\ref{Fig2}(b) for $n=$ 3, 4, 5, and 6 are calculated by the MPS method.
When $\varepsilon_0=0$, the transmission decreases as $n$ increases, see the square black dot line in Fig.~\ref{Fig2}(c) whose horizontal axis is inversely proportional.
Such a transmission decrease should be attributed to the photon-photon correlation induced by the TLA in the JCE.
As $n\gg1$, the TLA in JCE is consistently excited during the scattering process and no longer significantly impacts the transmission, resulting in that the transmission is primarily determined by the cavity.

If neglecting the photon-photon correlation, that is, assuming that only the firstly-arrived photon is scattered by the JCE while all the rest are scattered by the cavity, we can deduce that the transmission of an $n$-photon pulse equals $[T_J+(n-1)T_c]/n=T_c+(T_J-T_c)/n$, where $T_c$ and $T_J$ are the transmissions of the single photon state (i.e., $|i\rangle_1$) scattered by the OC and JCE, respectively.
The expression of $T_c+(T_J-T_c)/n$ is plotted as the circle red dotted line in Fig.~\ref{Fig2}(c).
Both lines converge toward $T_c$ as $n\rightarrow\infty$, see the grey dot.
Because the photon-photon correlation is fully neglected in  the red circle dot line, the difference between the two lines in Fig.~\ref{Fig2}(c), illustrated in Fig.~\ref{Fig2}(d), naturally reflects the strength of the photon-photon correlation induced by the JCE.
The line in Fig.~\ref{Fig2}(d) tells two points: 
the photon-photon correlation exerts the maximum influence as $n=3$, 
and the effect of the JCE as $n\rightarrow\infty$ resembles that of an OC, since the transmission tends to $T_c$.
These two conclusions will be further affirmed by the statistics analysis on the scattered states in Fig.~\ref{Fig3}.

{\sl Statistics.---}Figure~\ref{Fig3} presents the statistics of the final states $|f\rangle_n$ for the incident states $|i\rangle_n$ ($n=1,2,10$) scattered by the OC (a-e), TLA (f-j), and JCE (k-o).
The leftmost column shows the normalized photon-number distributions, i.e., $n_l/n$ with $n_l={}_{n} \langle f|\hat{a_l}^\dag\hat{a_l}|f\rangle_n$.
Note that the distributions $n_l/n$ for all $|i\rangle_n$ used are identical, see the gray curve in Fig.~\ref{Fig3}(a).
For the OC, the distributions $n_l/n$ in $|f\rangle_n$ are independent of $n$, see the black curve in Fig.~\ref{Fig3}(a).
However, for the TLA and JCE, the distributions $n_l/n$ show strong dependence on $n$, see Figs.~\ref{Fig3}(f) and \ref{Fig3}(k).
This is due to the single-excitation property of the TLA which only absorbs one photon one time.
Such a single excitation is responsible for the increase of the transmission in Fig.~\ref{Fig3}(f) as $n$ increases.
Since there is no photon-photon correlation in a single-photon state, the distribution $n_l/n$ for the TLA is identical to that for the OC when $n=1$, comparing the black lines in Figs.~\ref{Fig3}(f) and \ref{Fig3}(a). 
Being contrary to the TLA case, the transmission decreases as $n$ increases in the JCE one, see Fig.~\ref{Fig3}(k). 
This is due to that a JCE is constructed by an OC and a TLA.
Once the TLA is excited in multi-photon scattering process, the OC is responsible for the photon scattering.
Accordingly, the scattered state in the JCE case holds a similar distribution $n_l/n$ with that in the OC case for $n=10$, comparing the dotted red line in Fig.~\ref{Fig3}(k) with the solid black line in Fig.~\ref{Fig3}(a).
This confirms again that the tuning effect of the JCE on the incident photon states with $n\rightarrow\infty$ resembles that of an OC.

The dependence of $n_l/n$ on $n$ reflects the photon-photon correlation induced by the TLA or JCE, measured by the normalized second order correlation $g^{(2)}_{l}={}_{n}\! \left\langle f\right|\hat{a}^{\dag}_{l}\hat{a} ^{\dag}_{l}\hat{a}_{l}\hat{a}_{l}\left|f\right\rangle_{n}/n^2_l$ \cite{SI}, see Figs.~\ref{Fig3}(b, g, l).
Since $|i\rangle_n$ is a Fock state, we have its $g^{(2)}_{l}=1-{1\over n}$. 
In Fig.~\ref{Fig3}(b), the final and initial states have the identical $g^{(2)}_{l}$, which verifies that the OC neither adjusts the photon statistics nor results in a photon-photon correlation.
The circumstances change for the TLA and JCE due to the single-excitation property of the TLA, see Figs.~\ref{Fig3}(g) and \ref{Fig3}(l).
In Fig.~\ref{Fig3}(g), when $n=2$ (solid blue line) the transmitted and reflected photons are bunching ($g_l^{(2)}>1$) and anti-bunching ($g_l^{(2)}>1$), respectively, ascribed to the stimulated and spontaneous emission of the TLA \cite{shen2015photonic}.
If $n\rightarrow\infty$, the single-excitation property of the TLA would result in most of the incident photons are transmitted without feeling the TLA.
Physically speaking, as $n\rightarrow\infty$, $g_l^{(2)}$ for the transmitted pulse should gradually approach 1, while for the reflected pulse would become larger than 1, see the dashed red line in Fig.~\ref{Fig3}(g).

Different from the TLA, the JCE leads to that the transmitted and reflected photons exhibit anti-bunching and bunching behaviors for the bi-photon case, respectively, see Fig.~\ref{Fig3}(l).
As a composite of the OC and TLA, the JCE can be either singly or doubly excited at a time.
Since the TLA no longer strongly influence the incident states as $n\rightarrow\infty$, $g_l^{(2)}$ for both transmitted and reflected pulses approaches 1, see the dashed red line in Fig.~\ref{Fig3}(l).
This again confirms that it is the OC rather than the TLA in the JCE that predominantly governs the scattering process when $n\rightarrow\infty$, different from the few-photon situations.

To observe the probability of the state $|\sigma_l\rangle$ on each site, i.e., $\rho_l(\sigma_l)\equiv|\langle\sigma_l|f\rangle_n |^2$, we  plot them in the third and forth columns of Fig.~\ref{Fig3}. 
When $n=2$, the probability of $|2\rangle$ is always far less than that of $|1\rangle$ for all the OC, TLA, and JCE cases, even as $g^{(2)}_l>1$. 
This is due to that the probability of $|0\rangle$ approaches 1, leading to $\rho_l(1)\ll 1$.
Using $\rho_l(\sigma_l)$, we have $g_l^{(2)}={2\rho_l(2)\over[\rho_l(1) + 2\rho_l(2)]^2}$.
Since both $\rho_l(1)$ and $\rho_l(2)$ are far less than 1, $g_l^{(2)}>1$ as long as $\rho_l(2)\gtrsim0.5\rho_l^2(1)$ and subsequently, $\rho_l(2)\ll \rho_l(1) \ll 1$.
As $n$ increases, the probabilities of $|\sigma_l{\geqslant}2\rangle$ increase remarkably, see the forth column of Fig.~\ref{Fig3}, commonly holding $\rho_l(n)< \rho_l(n-1)$.

To measure the total contribution of $|\sigma_l{\geqslant}2\rangle$, we introduce the quantity $\eta_l\equiv1-\frac{\rho_l(1)}{n_l}$ to describe the probability of the photons coming from $|\sigma_l{\geqslant}2\rangle$, measuring the proportion of the bunching photons.
It is plotted in the last column of Fig.~\ref{Fig3} for $n=2$ and 10.
$\eta_l$ is very small as $n=2$ while become considerable as $n=10$, coinciding with the increase of $g_l^{(2)}$ as $n$ increases.

{\sl Conclusions.---}The MPS theory for the waveguide-QED systems is established to identify the novel statistics of the tens-of-photon states scattered by OCs, TLAs, and JCEs.
The OCs do not change the statistics of the photon states, while the TLA and JCE can tune the statistics of the scattered states and generate extremely different statistics between the 10-photon and bi-photon cases.
The photon-photon correlation induced by the JCE has a maximum strength when several photons are involved.
As the photon number tends to infinity, the tuning effect of the JCE on the incident states acts like the OC.
These distinguishable conclusions and the established matrix-product-state theory hold potential for the multi-photon operation in quantum informatics and techniques.


\begin{thebibliography}{48}%
\makeatletter
\providecommand \@ifxundefined [1]{%
 \@ifx{#1\undefined}
}%
\providecommand \@ifnum [1]{%
 \ifnum #1\expandafter \@firstoftwo
 \else \expandafter \@secondoftwo
 \fi
}%
\providecommand \@ifx [1]{%
 \ifx #1\expandafter \@firstoftwo
 \else \expandafter \@secondoftwo
 \fi
}%
\providecommand \natexlab [1]{#1}%
\providecommand \enquote  [1]{``#1''}%
\providecommand \bibnamefont  [1]{#1}%
\providecommand \bibfnamefont [1]{#1}%
\providecommand \citenamefont [1]{#1}%
\providecommand \href@noop [0]{\@secondoftwo}%
\providecommand \href [0]{\begingroup \@sanitize@url \@href}%
\providecommand \@href[1]{\@@startlink{#1}\@@href}%
\providecommand \@@href[1]{\endgroup#1\@@endlink}%
\providecommand \@sanitize@url [0]{\catcode `\\12\catcode `\$12\catcode
  `\&12\catcode `\#12\catcode `\^12\catcode `\_12\catcode `\%12\relax}%
\providecommand \@@startlink[1]{}%
\providecommand \@@endlink[0]{}%
\providecommand \url  [0]{\begingroup\@sanitize@url \@url }%
\providecommand \@url [1]{\endgroup\@href {#1}{\urlprefix }}%
\providecommand \urlprefix  [0]{URL }%
\providecommand \Eprint [0]{\href }%
\providecommand \doibase [0]{https://doi.org/}%
\providecommand \selectlanguage [0]{\@gobble}%
\providecommand \bibinfo  [0]{\@secondoftwo}%
\providecommand \bibfield  [0]{\@secondoftwo}%
\providecommand \translation [1]{[#1]}%
\providecommand \BibitemOpen [0]{}%
\providecommand \bibitemStop [0]{}%
\providecommand \bibitemNoStop [0]{.\EOS\space}%
\providecommand \EOS [0]{\spacefactor3000\relax}%
\providecommand \BibitemShut  [1]{\csname bibitem#1\endcsname}%
\let\auto@bib@innerbib\@empty
\bibitem [{\citenamefont {Vidal}(2004)}]{vidal2004efficient}%
  \BibitemOpen
  \bibfield  {author} {\bibinfo {author} {\bibfnamefont {G.}~\bibnamefont
  {Vidal}},\ }\bibfield  {title} {\bibinfo {title} {Efficient simulation of
  one-dimensional quantum many-body systems},\ }\href
  {https://doi.org/10.1103/PhysRevLett.93.040502} {\bibfield  {journal}
  {\bibinfo  {journal} {Phys. Rev. Lett.}\ }\textbf {\bibinfo {volume} {93}},\
  \bibinfo {pages} {040502} (\bibinfo {year} {2004})}\BibitemShut {NoStop}%
\bibitem [{\citenamefont {Eisert}\ \emph {et~al.}(2015)\citenamefont {Eisert},
  \citenamefont {Friesdorf},\ and\ \citenamefont
  {Gogolin}}]{eisert2015quantum}%
  \BibitemOpen
  \bibfield  {author} {\bibinfo {author} {\bibfnamefont {J.}~\bibnamefont
  {Eisert}}, \bibinfo {author} {\bibfnamefont {M.}~\bibnamefont {Friesdorf}},\
  and\ \bibinfo {author} {\bibfnamefont {C.}~\bibnamefont {Gogolin}},\
  }\bibfield  {title} {\bibinfo {title} {Quantum many-body systems out of
  equilibrium},\ }\href {https://doi.org/10.1038/nphys3215} {\bibfield
  {journal} {\bibinfo  {journal} {Nat. Phys.}\ }\textbf {\bibinfo {volume}
  {11}},\ \bibinfo {pages} {124} (\bibinfo {year} {2015})}\BibitemShut
  {NoStop}%
\bibitem [{\citenamefont {Carleo}\ and\ \citenamefont
  {Troyer}(2017)}]{carleo2017solving}%
  \BibitemOpen
  \bibfield  {author} {\bibinfo {author} {\bibfnamefont {G.}~\bibnamefont
  {Carleo}}\ and\ \bibinfo {author} {\bibfnamefont {M.}~\bibnamefont
  {Troyer}},\ }\bibfield  {title} {\bibinfo {title} {Solving the quantum
  many-body problem with artificial neural networks},\ }\href
  {https://doi.org/10.1103/PhysRevLett.130.100401} {\bibfield  {journal}
  {\bibinfo  {journal} {Science}\ }\textbf {\bibinfo {volume} {355}},\ \bibinfo
  {pages} {602} (\bibinfo {year} {2017})}\BibitemShut {NoStop}%
\bibitem [{\citenamefont {Weimer}\ \emph {et~al.}(2021)\citenamefont {Weimer},
  \citenamefont {Kshetrimayum},\ and\ \citenamefont
  {Or\'us}}]{weimer2021simulation}%
  \BibitemOpen
  \bibfield  {author} {\bibinfo {author} {\bibfnamefont {H.}~\bibnamefont
  {Weimer}}, \bibinfo {author} {\bibfnamefont {A.}~\bibnamefont
  {Kshetrimayum}},\ and\ \bibinfo {author} {\bibfnamefont {R.}~\bibnamefont
  {Or\'us}},\ }\bibfield  {title} {\bibinfo {title} {Simulation methods for
  open quantum many-body systems},\ }\href
  {https://doi.org/10.1103/RevModPhys.93.015008} {\bibfield  {journal}
  {\bibinfo  {journal} {Rev. Mod. Phys.}\ }\textbf {\bibinfo {volume} {93}},\
  \bibinfo {pages} {015008} (\bibinfo {year} {2021})}\BibitemShut {NoStop}%
\bibitem [{\citenamefont {Felser}\ \emph {et~al.}(2021)\citenamefont {Felser},
  \citenamefont {Notarnicola},\ and\ \citenamefont
  {Montangero}}]{felser2021efficient}%
  \BibitemOpen
  \bibfield  {author} {\bibinfo {author} {\bibfnamefont {T.}~\bibnamefont
  {Felser}}, \bibinfo {author} {\bibfnamefont {S.}~\bibnamefont
  {Notarnicola}},\ and\ \bibinfo {author} {\bibfnamefont {S.}~\bibnamefont
  {Montangero}},\ }\bibfield  {title} {\bibinfo {title} {Efficient tensor
  network ansatz for high-dimensional quantum many-body problems},\ }\href
  {https://doi.org/10.1103/PhysRevLett.126.170603} {\bibfield  {journal}
  {\bibinfo  {journal} {Phys. Rev. Lett.}\ }\textbf {\bibinfo {volume} {126}},\
  \bibinfo {pages} {170603} (\bibinfo {year} {2021})}\BibitemShut {NoStop}%
\bibitem [{\citenamefont {Shen}\ and\ \citenamefont
  {Fan}(2007)}]{shen2007strongly}%
  \BibitemOpen
  \bibfield  {author} {\bibinfo {author} {\bibfnamefont {J.-T.}\ \bibnamefont
  {Shen}}\ and\ \bibinfo {author} {\bibfnamefont {S.}~\bibnamefont {Fan}},\
  }\bibfield  {title} {\bibinfo {title} {Strongly correlated two-photon
  transport in a one-dimensional waveguide coupled to a two-level system},\
  }\href {https://doi.org/10.1103/PhysRevLett.98.153003} {\bibfield  {journal}
  {\bibinfo  {journal} {Phys. Rev. Lett.}\ }\textbf {\bibinfo {volume} {98}},\
  \bibinfo {pages} {153003} (\bibinfo {year} {2007})}\BibitemShut {NoStop}%
\bibitem [{\citenamefont {Yudson}\ and\ \citenamefont
  {Reineker}(2008)}]{yudson2008multiphoton}%
  \BibitemOpen
  \bibfield  {author} {\bibinfo {author} {\bibfnamefont {V.~I.}\ \bibnamefont
  {Yudson}}\ and\ \bibinfo {author} {\bibfnamefont {P.}~\bibnamefont
  {Reineker}},\ }\bibfield  {title} {\bibinfo {title} {Multiphoton scattering
  in a one-dimensional waveguide with resonant atoms},\ }\href
  {https://doi.org/10.1103/PhysRevA.78.052713} {\bibfield  {journal} {\bibinfo
  {journal} {Phys. Rev. A}\ }\textbf {\bibinfo {volume} {78}},\ \bibinfo
  {pages} {052713} (\bibinfo {year} {2008})}\BibitemShut {NoStop}%
\bibitem [{\citenamefont {Pan}\ \emph {et~al.}(2012)\citenamefont {Pan},
  \citenamefont {Chen}, \citenamefont {Lu}, \citenamefont {Weinfurter},
  \citenamefont {Zeilinger},\ and\ \citenamefont {\ifmmode~\dot{Z}\else
  \.{Z}\fi{}ukowski}}]{pan2012multiphoton}%
  \BibitemOpen
  \bibfield  {author} {\bibinfo {author} {\bibfnamefont {J.-W.}\ \bibnamefont
  {Pan}}, \bibinfo {author} {\bibfnamefont {Z.-B.}\ \bibnamefont {Chen}},
  \bibinfo {author} {\bibfnamefont {C.-Y.}\ \bibnamefont {Lu}}, \bibinfo
  {author} {\bibfnamefont {H.}~\bibnamefont {Weinfurter}}, \bibinfo {author}
  {\bibfnamefont {A.}~\bibnamefont {Zeilinger}},\ and\ \bibinfo {author}
  {\bibfnamefont {M.}~\bibnamefont {\ifmmode~\dot{Z}\else \.{Z}\fi{}ukowski}},\
  }\bibfield  {title} {\bibinfo {title} {Multiphoton entanglement and
  interferometry},\ }\href {https://doi.org/10.1103/RevModPhys.84.777}
  {\bibfield  {journal} {\bibinfo  {journal} {Rev. Mod. Phys.}\ }\textbf
  {\bibinfo {volume} {84}},\ \bibinfo {pages} {777} (\bibinfo {year}
  {2012})}\BibitemShut {NoStop}%
\bibitem [{\citenamefont {Baragiola}\ \emph {et~al.}(2012)\citenamefont
  {Baragiola}, \citenamefont {Cook}, \citenamefont {Bra\ifmmode~\acute{n}\else
  \'{n}\fi{}czyk},\ and\ \citenamefont {Combes}}]{baragiola2012n}%
  \BibitemOpen
  \bibfield  {author} {\bibinfo {author} {\bibfnamefont {B.~Q.}\ \bibnamefont
  {Baragiola}}, \bibinfo {author} {\bibfnamefont {R.~L.}\ \bibnamefont {Cook}},
  \bibinfo {author} {\bibfnamefont {A.~M.}\ \bibnamefont
  {Bra\ifmmode~\acute{n}\else \'{n}\fi{}czyk}},\ and\ \bibinfo {author}
  {\bibfnamefont {J.}~\bibnamefont {Combes}},\ }\bibfield  {title} {\bibinfo
  {title} {$n$-photon wave packets interacting with an arbitrary quantum
  system},\ }\href {https://doi.org/10.1103/PhysRevA.86.013811} {\bibfield
  {journal} {\bibinfo  {journal} {Phys. Rev. A}\ }\textbf {\bibinfo {volume}
  {86}},\ \bibinfo {pages} {013811} (\bibinfo {year} {2012})}\BibitemShut
  {NoStop}%
\bibitem [{\citenamefont {Shi}\ \emph {et~al.}(2015)\citenamefont {Shi},
  \citenamefont {Chang},\ and\ \citenamefont {Cirac}}]{shi2015multiphoton}%
  \BibitemOpen
  \bibfield  {author} {\bibinfo {author} {\bibfnamefont {T.}~\bibnamefont
  {Shi}}, \bibinfo {author} {\bibfnamefont {D.~E.}\ \bibnamefont {Chang}},\
  and\ \bibinfo {author} {\bibfnamefont {J.~I.}\ \bibnamefont {Cirac}},\
  }\bibfield  {title} {\bibinfo {title} {Multiphoton-scattering theory and
  generalized master equations},\ }\href
  {https://doi.org/10.1103/PhysRevA.92.053834} {\bibfield  {journal} {\bibinfo
  {journal} {Phys. Rev. A}\ }\textbf {\bibinfo {volume} {92}},\ \bibinfo
  {pages} {053834} (\bibinfo {year} {2015})}\BibitemShut {NoStop}%
\bibitem [{\citenamefont {Kocaba\ifmmode~\mbox{\c{s}}\else
  \c{s}\fi{}}(2016)}]{Kocaba2016Effects}%
  \BibitemOpen
  \bibfield  {author} {\bibinfo {author} {\bibfnamefont {i.~m. c.~E.}\
  \bibnamefont {Kocaba\ifmmode~\mbox{\c{s}}\else \c{s}\fi{}}},\ }\bibfield
  {title} {\bibinfo {title} {Effects of modal dispersion on few-photon--qubit
  scattering in one-dimensional waveguides},\ }\href
  {https://doi.org/10.1103/PhysRevA.93.033829} {\bibfield  {journal} {\bibinfo
  {journal} {Phys. Rev. A}\ }\textbf {\bibinfo {volume} {93}},\ \bibinfo
  {pages} {033829} (\bibinfo {year} {2016})}\BibitemShut {NoStop}%
\bibitem [{\citenamefont {Liao}\ \emph {et~al.}(2020)\citenamefont {Liao},
  \citenamefont {Lu},\ and\ \citenamefont {Zubairy}}]{liao2020multiphoton}%
  \BibitemOpen
  \bibfield  {author} {\bibinfo {author} {\bibfnamefont {Z.}~\bibnamefont
  {Liao}}, \bibinfo {author} {\bibfnamefont {Y.}~\bibnamefont {Lu}},\ and\
  \bibinfo {author} {\bibfnamefont {M.~S.}\ \bibnamefont {Zubairy}},\
  }\bibfield  {title} {\bibinfo {title} {Multiphoton pulses interacting with
  multiple emitters in a one-dimensional waveguide},\ }\href
  {https://doi.org/10.1103/PhysRevA.102.053702} {\bibfield  {journal} {\bibinfo
   {journal} {Phys. Rev. A}\ }\textbf {\bibinfo {volume} {102}},\ \bibinfo
  {pages} {053702} (\bibinfo {year} {2020})}\BibitemShut {NoStop}%
\bibitem [{\citenamefont {Zhong}\ \emph {et~al.}(2020)\citenamefont {Zhong},
  \citenamefont {Wang}, \citenamefont {Deng}, \citenamefont {Chen},
  \citenamefont {Peng}, \citenamefont {Luo}, \citenamefont {Qin}, \citenamefont
  {Wu}, \citenamefont {Ding}, \citenamefont {Hu}, \citenamefont {Hu},
  \citenamefont {Yang}, \citenamefont {Zhang}, \citenamefont {Li},
  \citenamefont {Li}, \citenamefont {Jiang}, \citenamefont {Gan}, \citenamefont
  {Yang}, \citenamefont {You}, \citenamefont {Wang}, \citenamefont {Li},
  \citenamefont {Liu}, \citenamefont {Lu},\ and\ \citenamefont
  {Pan}}]{zhong2020quantum}%
  \BibitemOpen
  \bibfield  {author} {\bibinfo {author} {\bibfnamefont {H.-S.}\ \bibnamefont
  {Zhong}}, \bibinfo {author} {\bibfnamefont {H.}~\bibnamefont {Wang}},
  \bibinfo {author} {\bibfnamefont {Y.-H.}\ \bibnamefont {Deng}}, \bibinfo
  {author} {\bibfnamefont {M.-C.}\ \bibnamefont {Chen}}, \bibinfo {author}
  {\bibfnamefont {L.-C.}\ \bibnamefont {Peng}}, \bibinfo {author}
  {\bibfnamefont {Y.-H.}\ \bibnamefont {Luo}}, \bibinfo {author} {\bibfnamefont
  {J.}~\bibnamefont {Qin}}, \bibinfo {author} {\bibfnamefont {D.}~\bibnamefont
  {Wu}}, \bibinfo {author} {\bibfnamefont {X.}~\bibnamefont {Ding}}, \bibinfo
  {author} {\bibfnamefont {Y.}~\bibnamefont {Hu}}, \bibinfo {author}
  {\bibfnamefont {P.}~\bibnamefont {Hu}}, \bibinfo {author} {\bibfnamefont
  {X.-Y.}\ \bibnamefont {Yang}}, \bibinfo {author} {\bibfnamefont {W.-J.}\
  \bibnamefont {Zhang}}, \bibinfo {author} {\bibfnamefont {H.}~\bibnamefont
  {Li}}, \bibinfo {author} {\bibfnamefont {Y.}~\bibnamefont {Li}}, \bibinfo
  {author} {\bibfnamefont {X.}~\bibnamefont {Jiang}}, \bibinfo {author}
  {\bibfnamefont {L.}~\bibnamefont {Gan}}, \bibinfo {author} {\bibfnamefont
  {G.}~\bibnamefont {Yang}}, \bibinfo {author} {\bibfnamefont {L.}~\bibnamefont
  {You}}, \bibinfo {author} {\bibfnamefont {Z.}~\bibnamefont {Wang}}, \bibinfo
  {author} {\bibfnamefont {L.}~\bibnamefont {Li}}, \bibinfo {author}
  {\bibfnamefont {N.-L.}\ \bibnamefont {Liu}}, \bibinfo {author} {\bibfnamefont
  {C.-Y.}\ \bibnamefont {Lu}},\ and\ \bibinfo {author} {\bibfnamefont {J.-W.}\
  \bibnamefont {Pan}},\ }\bibfield  {title} {\bibinfo {title} {Quantum
  computational advantage using photons},\ }\href
  {https://www.science.org/doi/full/10.1126/science.abe8770} {\bibfield
  {journal} {\bibinfo  {journal} {Science}\ }\textbf {\bibinfo {volume}
  {370}},\ \bibinfo {pages} {1460} (\bibinfo {year} {2020})}\BibitemShut
  {NoStop}%
\bibitem [{\citenamefont {Shen}\ and\ \citenamefont
  {Fan}(2009)}]{shen2009theory}%
  \BibitemOpen
  \bibfield  {author} {\bibinfo {author} {\bibfnamefont {J.-T.}\ \bibnamefont
  {Shen}}\ and\ \bibinfo {author} {\bibfnamefont {S.}~\bibnamefont {Fan}},\
  }\bibfield  {title} {\bibinfo {title} {Theory of single-photon transport in a
  single-mode waveguide. i. coupling to a cavity containing a two-level atom},\
  }\href {https://doi.org/10.1103/PhysRevA.79.023837} {\bibfield  {journal}
  {\bibinfo  {journal} {Phys. Rev. A}\ }\textbf {\bibinfo {volume} {79}},\
  \bibinfo {pages} {023837} (\bibinfo {year} {2009})}\BibitemShut {NoStop}%
\bibitem [{\citenamefont {Liao}\ \emph {et~al.}(2016)\citenamefont {Liao},
  \citenamefont {Zeng}, \citenamefont {Nha},\ and\ \citenamefont
  {Zubairy}}]{Liao_2016}%
  \BibitemOpen
  \bibfield  {author} {\bibinfo {author} {\bibfnamefont {Z.}~\bibnamefont
  {Liao}}, \bibinfo {author} {\bibfnamefont {X.}~\bibnamefont {Zeng}}, \bibinfo
  {author} {\bibfnamefont {H.}~\bibnamefont {Nha}},\ and\ \bibinfo {author}
  {\bibfnamefont {M.~S.}\ \bibnamefont {Zubairy}},\ }\bibfield  {title}
  {\bibinfo {title} {Photon transport in a one-dimensional nanophotonic
  waveguide qed system},\ }\href
  {https://doi.org/10.1088/0031-8949/91/6/063004} {\bibfield  {journal}
  {\bibinfo  {journal} {Phys. Scr.}\ }\textbf {\bibinfo {volume} {91}},\
  \bibinfo {pages} {063004} (\bibinfo {year} {2016})}\BibitemShut {NoStop}%
\bibitem [{\citenamefont {Sheremet}\ \emph {et~al.}(2023)\citenamefont
  {Sheremet}, \citenamefont {Petrov}, \citenamefont {Iorsh}, \citenamefont
  {Poshakinskiy},\ and\ \citenamefont {Poddubny}}]{sheremet2023waveguide}%
  \BibitemOpen
  \bibfield  {author} {\bibinfo {author} {\bibfnamefont {A.~S.}\ \bibnamefont
  {Sheremet}}, \bibinfo {author} {\bibfnamefont {M.~I.}\ \bibnamefont
  {Petrov}}, \bibinfo {author} {\bibfnamefont {I.~V.}\ \bibnamefont {Iorsh}},
  \bibinfo {author} {\bibfnamefont {A.~V.}\ \bibnamefont {Poshakinskiy}},\ and\
  \bibinfo {author} {\bibfnamefont {A.~N.}\ \bibnamefont {Poddubny}},\
  }\bibfield  {title} {\bibinfo {title} {Waveguide quantum electrodynamics:
  Collective radiance and photon-photon correlations},\ }\href
  {https://doi.org/10.1103/RevModPhys.95.015002} {\bibfield  {journal}
  {\bibinfo  {journal} {Rev. Mod. Phys.}\ }\textbf {\bibinfo {volume} {95}},\
  \bibinfo {pages} {015002} (\bibinfo {year} {2023})}\BibitemShut {NoStop}%
\bibitem [{\citenamefont {Waks}\ and\ \citenamefont
  {Vuckovic}(2006)}]{waks2006dipole}%
  \BibitemOpen
  \bibfield  {author} {\bibinfo {author} {\bibfnamefont {E.}~\bibnamefont
  {Waks}}\ and\ \bibinfo {author} {\bibfnamefont {J.}~\bibnamefont
  {Vuckovic}},\ }\bibfield  {title} {\bibinfo {title} {Dipole induced
  transparency in drop-filter cavity-waveguide systems},\ }\href
  {https://doi.org/10.1103/PhysRevLett.96.153601} {\bibfield  {journal}
  {\bibinfo  {journal} {Phys. Rev. Lett.}\ }\textbf {\bibinfo {volume} {96}},\
  \bibinfo {pages} {153601} (\bibinfo {year} {2006})}\BibitemShut {NoStop}%
\bibitem [{\citenamefont {Shi}\ \emph {et~al.}(2011)\citenamefont {Shi},
  \citenamefont {Fan},\ and\ \citenamefont {Sun}}]{shi2011two}%
  \BibitemOpen
  \bibfield  {author} {\bibinfo {author} {\bibfnamefont {T.}~\bibnamefont
  {Shi}}, \bibinfo {author} {\bibfnamefont {S.}~\bibnamefont {Fan}},\ and\
  \bibinfo {author} {\bibfnamefont {C.~P.}\ \bibnamefont {Sun}},\ }\bibfield
  {title} {\bibinfo {title} {Two-photon transport in a waveguide coupled to a
  cavity in a two-level system},\ }\href
  {https://doi.org/10.1103/PhysRevA.84.063803} {\bibfield  {journal} {\bibinfo
  {journal} {Phys. Rev. A}\ }\textbf {\bibinfo {volume} {84}},\ \bibinfo
  {pages} {063803} (\bibinfo {year} {2011})}\BibitemShut {NoStop}%
\bibitem [{\citenamefont {Li}\ \emph {et~al.}(2023)\citenamefont {Li},
  \citenamefont {Li}, \citenamefont {Zeng}, \citenamefont {Hu}, \citenamefont
  {Xu}, \citenamefont {Zhou}, \citenamefont {Xia}, \citenamefont {Xu},\ and\
  \citenamefont {Yang}}]{li2023single}%
  \BibitemOpen
  \bibfield  {author} {\bibinfo {author} {\bibfnamefont {H.}~\bibnamefont
  {Li}}, \bibinfo {author} {\bibfnamefont {Z.}~\bibnamefont {Li}}, \bibinfo
  {author} {\bibfnamefont {R.}~\bibnamefont {Zeng}}, \bibinfo {author}
  {\bibfnamefont {M.}~\bibnamefont {Hu}}, \bibinfo {author} {\bibfnamefont
  {M.}~\bibnamefont {Xu}}, \bibinfo {author} {\bibfnamefont {X.}~\bibnamefont
  {Zhou}}, \bibinfo {author} {\bibfnamefont {X.}~\bibnamefont {Xia}}, \bibinfo
  {author} {\bibfnamefont {J.}~\bibnamefont {Xu}},\ and\ \bibinfo {author}
  {\bibfnamefont {Y.}~\bibnamefont {Yang}},\ }\bibfield  {title} {\bibinfo
  {title} {Single-photon switching in a floquet waveguide-qed system with
  time-modulated coupling constants},\ }\href
  {https://doi.org/10.1103/PhysRevA.107.023720} {\bibfield  {journal} {\bibinfo
   {journal} {Phys. Rev. A}\ }\textbf {\bibinfo {volume} {107}},\ \bibinfo
  {pages} {023720} (\bibinfo {year} {2023})}\BibitemShut {NoStop}%
\bibitem [{\citenamefont {Rephaeli}\ \emph {et~al.}(2011)\citenamefont
  {Rephaeli}, \citenamefont {Kocaba{\c s}},\ and\ \citenamefont
  {Fan}}]{rephaeli2011few}%
  \BibitemOpen
  \bibfield  {author} {\bibinfo {author} {\bibfnamefont {E.}~\bibnamefont
  {Rephaeli}}, \bibinfo {author} {\bibfnamefont {{\c S}.~E.}\ \bibnamefont
  {Kocaba{\c s}}},\ and\ \bibinfo {author} {\bibfnamefont {S.}~\bibnamefont
  {Fan}},\ }\bibfield  {title} {\bibinfo {title} {Few-photon transport in a
  waveguide coupled to a pair of colocated two-level atoms},\ }\href
  {https://doi.org/10.1103/PhysRevA.84.063832} {\bibfield  {journal} {\bibinfo
  {journal} {Phys. Rev. A}\ }\textbf {\bibinfo {volume} {84}},\ \bibinfo
  {pages} {063832} (\bibinfo {year} {2011})}\BibitemShut {NoStop}%
\bibitem [{\citenamefont {Hu}\ \emph {et~al.}(2018)\citenamefont {Hu},
  \citenamefont {Zou},\ and\ \citenamefont {Zhang}}]{Hu2018Transmission}%
  \BibitemOpen
  \bibfield  {author} {\bibinfo {author} {\bibfnamefont {Q.}~\bibnamefont
  {Hu}}, \bibinfo {author} {\bibfnamefont {B.}~\bibnamefont {Zou}},\ and\
  \bibinfo {author} {\bibfnamefont {Y.}~\bibnamefont {Zhang}},\ }\bibfield
  {title} {\bibinfo {title} {Transmission and correlation of a two-photon pulse
  in a one-dimensional waveguide coupled with quantum emitters},\ }\href
  {https://doi.org/10.1103/PhysRevA.97.033847} {\bibfield  {journal} {\bibinfo
  {journal} {Phys. Rev. A}\ }\textbf {\bibinfo {volume} {97}},\ \bibinfo
  {pages} {033847} (\bibinfo {year} {2018})}\BibitemShut {NoStop}%
\bibitem [{\citenamefont {Hofheinz}\ \emph {et~al.}(2008)\citenamefont
  {Hofheinz}, \citenamefont {Weig}, \citenamefont {Ansmann}, \citenamefont
  {Bialczak},\ and\ \citenamefont {Cleland}}]{2008Generation}%
  \BibitemOpen
  \bibfield  {author} {\bibinfo {author} {\bibfnamefont {M.}~\bibnamefont
  {Hofheinz}}, \bibinfo {author} {\bibfnamefont {E.~M.}\ \bibnamefont {Weig}},
  \bibinfo {author} {\bibfnamefont {M.}~\bibnamefont {Ansmann}}, \bibinfo
  {author} {\bibfnamefont {R.~C.}\ \bibnamefont {Bialczak}},\ and\ \bibinfo
  {author} {\bibfnamefont {A.~N.}\ \bibnamefont {Cleland}},\ }\bibfield
  {title} {\bibinfo {title} {Generation of fock states in a superconducting
  quantum circuit},\ }\href {https://doi.org/10.1038/nature07136} {\bibfield
  {journal} {\bibinfo  {journal} {Nature}\ }\textbf {\bibinfo {volume} {454}},\
  \bibinfo {pages} {310} (\bibinfo {year} {2008})}\BibitemShut {NoStop}%
\bibitem [{\citenamefont {Matthews}\ \emph {et~al.}(2009)\citenamefont
  {Matthews}, \citenamefont {Politi}, \citenamefont {Stefanov},\ and\
  \citenamefont {O'brien}}]{matthews2009manipulation}%
  \BibitemOpen
  \bibfield  {author} {\bibinfo {author} {\bibfnamefont {J.~C.}\ \bibnamefont
  {Matthews}}, \bibinfo {author} {\bibfnamefont {A.}~\bibnamefont {Politi}},
  \bibinfo {author} {\bibfnamefont {A.}~\bibnamefont {Stefanov}},\ and\
  \bibinfo {author} {\bibfnamefont {J.~L.}\ \bibnamefont {O'brien}},\
  }\bibfield  {title} {\bibinfo {title} {Manipulation of multiphoton
  entanglement in waveguide quantum circuits},\ }\href
  {https://doi.org/10.1038/nphoton.2009.93} {\bibfield  {journal} {\bibinfo
  {journal} {Nat. Photonics}\ }\textbf {\bibinfo {volume} {3}},\ \bibinfo
  {pages} {346} (\bibinfo {year} {2009})}\BibitemShut {NoStop}%
\bibitem [{\citenamefont {Shen}\ and\ \citenamefont
  {Shen}(2015)}]{shen2015photonic}%
  \BibitemOpen
  \bibfield  {author} {\bibinfo {author} {\bibfnamefont {Y.}~\bibnamefont
  {Shen}}\ and\ \bibinfo {author} {\bibfnamefont {J.-T.}\ \bibnamefont
  {Shen}},\ }\bibfield  {title} {\bibinfo {title} {Photonic-fock-state
  scattering in a waveguide-qed system and their correlation functions},\
  }\href {https://doi.org/10.1103/PhysRevA.92.033803} {\bibfield  {journal}
  {\bibinfo  {journal} {Phys. Rev. A}\ }\textbf {\bibinfo {volume} {92}},\
  \bibinfo {pages} {033803} (\bibinfo {year} {2015})}\BibitemShut {NoStop}%
\bibitem [{\citenamefont {Mirza}\ and\ \citenamefont
  {Schotland}(2016)}]{Miraz2016Two}%
  \BibitemOpen
  \bibfield  {author} {\bibinfo {author} {\bibfnamefont {I.~M.}\ \bibnamefont
  {Mirza}}\ and\ \bibinfo {author} {\bibfnamefont {J.~C.}\ \bibnamefont
  {Schotland}},\ }\bibfield  {title} {\bibinfo {title} {Two-photon entanglement
  in multiqubit bidirectional-waveguide qed},\ }\href
  {https://doi.org/10.1103/PhysRevA.94.012309} {\bibfield  {journal} {\bibinfo
  {journal} {Phys. Rev. A}\ }\textbf {\bibinfo {volume} {94}},\ \bibinfo
  {pages} {012309} (\bibinfo {year} {2016})}\BibitemShut {NoStop}%
\bibitem [{\citenamefont {Zheng}\ \emph {et~al.}(2010)\citenamefont {Zheng},
  \citenamefont {Gauthier},\ and\ \citenamefont
  {Baranger}}]{zheng2010waveguide}%
  \BibitemOpen
  \bibfield  {author} {\bibinfo {author} {\bibfnamefont {H.}~\bibnamefont
  {Zheng}}, \bibinfo {author} {\bibfnamefont {D.~J.}\ \bibnamefont
  {Gauthier}},\ and\ \bibinfo {author} {\bibfnamefont {H.~U.}\ \bibnamefont
  {Baranger}},\ }\bibfield  {title} {\bibinfo {title} {Waveguide qed: Many-body
  bound-state effects in coherent and fock-state scattering from a two-level
  system},\ }\href {https://doi.org/10.1103/PhysRevA.82.063816} {\bibfield
  {journal} {\bibinfo  {journal} {Phys. Rev. A}\ }\textbf {\bibinfo {volume}
  {82}},\ \bibinfo {pages} {063816} (\bibinfo {year} {2010})}\BibitemShut
  {NoStop}%
\bibitem [{\citenamefont {Shi}\ and\ \citenamefont
  {Sun}(2009)}]{PhysRevB.79.205111}%
  \BibitemOpen
  \bibfield  {author} {\bibinfo {author} {\bibfnamefont {T.}~\bibnamefont
  {Shi}}\ and\ \bibinfo {author} {\bibfnamefont {C.~P.}\ \bibnamefont {Sun}},\
  }\bibfield  {title} {\bibinfo {title} {Lehmann-symanzik-zimmermann reduction
  approach to multiphoton scattering in coupled-resonator arrays},\ }\href
  {https://doi.org/10.1103/PhysRevB.79.205111} {\bibfield  {journal} {\bibinfo
  {journal} {Phys. Rev. B}\ }\textbf {\bibinfo {volume} {79}},\ \bibinfo
  {pages} {205111} (\bibinfo {year} {2009})}\BibitemShut {NoStop}%
\bibitem [{\citenamefont {Shi}\ and\ \citenamefont {Fan}(2013)}]{Shi2013Two}%
  \BibitemOpen
  \bibfield  {author} {\bibinfo {author} {\bibfnamefont {T.}~\bibnamefont
  {Shi}}\ and\ \bibinfo {author} {\bibfnamefont {S.}~\bibnamefont {Fan}},\
  }\bibfield  {title} {\bibinfo {title} {Two-photon transport through a
  waveguide coupling to a whispering-gallery resonator containing an atom and
  photon-blockade effect},\ }\href {https://doi.org/10.1103/PhysRevA.87.063818}
  {\bibfield  {journal} {\bibinfo  {journal} {Phys. Rev. A}\ }\textbf {\bibinfo
  {volume} {87}},\ \bibinfo {pages} {063818} (\bibinfo {year}
  {2013})}\BibitemShut {NoStop}%
\bibitem [{\citenamefont {See}\ \emph {et~al.}(2017)\citenamefont {See},
  \citenamefont {Noh},\ and\ \citenamefont {Angelakis}}]{see2017diagrammatic}%
  \BibitemOpen
  \bibfield  {author} {\bibinfo {author} {\bibfnamefont {T.~F.}\ \bibnamefont
  {See}}, \bibinfo {author} {\bibfnamefont {C.}~\bibnamefont {Noh}},\ and\
  \bibinfo {author} {\bibfnamefont {D.~G.}\ \bibnamefont {Angelakis}},\
  }\bibfield  {title} {\bibinfo {title} {Diagrammatic approach to multiphoton
  scattering},\ }\href {https://doi.org/10.1103/PhysRevA.95.053845} {\bibfield
  {journal} {\bibinfo  {journal} {Phys. Rev. A}\ }\textbf {\bibinfo {volume}
  {95}},\ \bibinfo {pages} {053845} (\bibinfo {year} {2017})}\BibitemShut
  {NoStop}%
\bibitem [{\citenamefont {Fan}\ \emph {et~al.}(2010)\citenamefont {Fan},
  \citenamefont {Kocaba\ifmmode~\mbox{\c{s}}\else \c{s}\fi{}},\ and\
  \citenamefont {Shen}}]{Fan2010Input}%
  \BibitemOpen
  \bibfield  {author} {\bibinfo {author} {\bibfnamefont {S.}~\bibnamefont
  {Fan}}, \bibinfo {author} {\bibfnamefont {i.~m. c.~E.}\ \bibnamefont
  {Kocaba\ifmmode~\mbox{\c{s}}\else \c{s}\fi{}}},\ and\ \bibinfo {author}
  {\bibfnamefont {J.-T.}\ \bibnamefont {Shen}},\ }\bibfield  {title} {\bibinfo
  {title} {Input-output formalism for few-photon transport in one-dimensional
  nanophotonic waveguides coupled to a qubit},\ }\href
  {https://doi.org/10.1103/PhysRevA.82.063821} {\bibfield  {journal} {\bibinfo
  {journal} {Phys. Rev. A}\ }\textbf {\bibinfo {volume} {82}},\ \bibinfo
  {pages} {063821} (\bibinfo {year} {2010})}\BibitemShut {NoStop}%
\bibitem [{\citenamefont {Rephaeli}\ and\ \citenamefont {Fan}(2012)}]{6189728}%
  \BibitemOpen
  \bibfield  {author} {\bibinfo {author} {\bibfnamefont {E.}~\bibnamefont
  {Rephaeli}}\ and\ \bibinfo {author} {\bibfnamefont {S.}~\bibnamefont {Fan}},\
  }\bibfield  {title} {\bibinfo {title} {Few-photon single-atom cavity qed with
  input-output formalism in fock space},\ }\href
  {https://doi.org/10.1109/JSTQE.2012.2196261} {\bibfield  {journal} {\bibinfo
  {journal} {IEEE J. Sel. Top. Quantum Electron.}\ }\textbf {\bibinfo {volume}
  {18}},\ \bibinfo {pages} {1754} (\bibinfo {year} {2012})}\BibitemShut
  {NoStop}%
\bibitem [{\citenamefont {Xu}\ and\ \citenamefont {Fan}(2015)}]{Xu2015Input}%
  \BibitemOpen
  \bibfield  {author} {\bibinfo {author} {\bibfnamefont {S.}~\bibnamefont
  {Xu}}\ and\ \bibinfo {author} {\bibfnamefont {S.}~\bibnamefont {Fan}},\
  }\bibfield  {title} {\bibinfo {title} {Input-output formalism for few-photon
  transport: A systematic treatment beyond two photons},\ }\href
  {https://doi.org/10.1103/PhysRevA.91.043845} {\bibfield  {journal} {\bibinfo
  {journal} {Phys. Rev. A}\ }\textbf {\bibinfo {volume} {91}},\ \bibinfo
  {pages} {043845} (\bibinfo {year} {2015})}\BibitemShut {NoStop}%
\bibitem [{\citenamefont {Joanesarson}\ \emph {et~al.}(2020)\citenamefont
  {Joanesarson}, \citenamefont {Iles-Smith}, \citenamefont {Heuck},\ and\
  \citenamefont {M\o{}rk}}]{PhysRevA.101.063809}%
  \BibitemOpen
  \bibfield  {author} {\bibinfo {author} {\bibfnamefont {K.~B.}\ \bibnamefont
  {Joanesarson}}, \bibinfo {author} {\bibfnamefont {J.}~\bibnamefont
  {Iles-Smith}}, \bibinfo {author} {\bibfnamefont {M.}~\bibnamefont {Heuck}},\
  and\ \bibinfo {author} {\bibfnamefont {J.}~\bibnamefont {M\o{}rk}},\
  }\bibfield  {title} {\bibinfo {title} {Few-photon transport in fano-resonance
  waveguide geometries},\ }\href {https://doi.org/10.1103/PhysRevA.101.063809}
  {\bibfield  {journal} {\bibinfo  {journal} {Phys. Rev. A}\ }\textbf {\bibinfo
  {volume} {101}},\ \bibinfo {pages} {063809} (\bibinfo {year}
  {2020})}\BibitemShut {NoStop}%
\bibitem [{\citenamefont {Anderson}(1972)}]{anderson1972more}%
  \BibitemOpen
  \bibfield  {author} {\bibinfo {author} {\bibfnamefont {P.~W.}\ \bibnamefont
  {Anderson}},\ }\bibfield  {title} {\bibinfo {title} {More is different:
  Broken symmetry and the nature of the hierarchical structure of science.},\
  }\href
  {https://www.science.org/doi/abs/10.1126/science.177.4047.393#tab-citations}
  {\bibfield  {journal} {\bibinfo  {journal} {Science}\ }\textbf {\bibinfo
  {volume} {177}},\ \bibinfo {pages} {393} (\bibinfo {year}
  {1972})}\BibitemShut {NoStop}%
\bibitem [{\citenamefont {Schollw\"{o}ck}(2011)}]{SCHOLLWOCK201196}%
  \BibitemOpen
  \bibfield  {author} {\bibinfo {author} {\bibfnamefont {U.}~\bibnamefont
  {Schollw\"{o}ck}},\ }\bibfield  {title} {\bibinfo {title} {The density-matrix
  renormalization group in the age of matrix product states},\ }\href
  {https://doi.org/https://doi.org/10.1016/j.aop.2010.09.012} {\bibfield
  {journal} {\bibinfo  {journal} {Annals of Physics}\ }\textbf {\bibinfo
  {volume} {326}},\ \bibinfo {pages} {96} (\bibinfo {year} {2011})},\ \bibinfo
  {note} {january 2011 Special Issue}\BibitemShut {NoStop}%
\bibitem [{\citenamefont {Paeckel}\ \emph {et~al.}(2019)\citenamefont
  {Paeckel}, \citenamefont {K\"{o}hler}, \citenamefont {Swoboda}, \citenamefont
  {Manmana}, \citenamefont {Schollw\"{o}ck},\ and\ \citenamefont
  {Hubig}}]{PAECKEL2019167998}%
  \BibitemOpen
  \bibfield  {author} {\bibinfo {author} {\bibfnamefont {S.}~\bibnamefont
  {Paeckel}}, \bibinfo {author} {\bibfnamefont {T.}~\bibnamefont {K\"{o}hler}},
  \bibinfo {author} {\bibfnamefont {A.}~\bibnamefont {Swoboda}}, \bibinfo
  {author} {\bibfnamefont {S.~R.}\ \bibnamefont {Manmana}}, \bibinfo {author}
  {\bibfnamefont {U.}~\bibnamefont {Schollw\"{o}ck}},\ and\ \bibinfo {author}
  {\bibfnamefont {C.}~\bibnamefont {Hubig}},\ }\bibfield  {title} {\bibinfo
  {title} {Time-evolution methods for matrix-product states},\ }\href
  {https://doi.org/https://doi.org/10.1016/j.aop.2019.167998} {\bibfield
  {journal} {\bibinfo  {journal} {Annals of Physics}\ }\textbf {\bibinfo
  {volume} {411}},\ \bibinfo {pages} {167998} (\bibinfo {year}
  {2019})}\BibitemShut {NoStop}%
\bibitem [{\citenamefont {Cirac}\ \emph {et~al.}(2021)\citenamefont {Cirac},
  \citenamefont {P\'erez-Garc\'{\i}a}, \citenamefont {Schuch},\ and\
  \citenamefont {Verstraete}}]{RevModPhys.93.045003}%
  \BibitemOpen
  \bibfield  {author} {\bibinfo {author} {\bibfnamefont {J.~I.}\ \bibnamefont
  {Cirac}}, \bibinfo {author} {\bibfnamefont {D.}~\bibnamefont
  {P\'erez-Garc\'{\i}a}}, \bibinfo {author} {\bibfnamefont {N.}~\bibnamefont
  {Schuch}},\ and\ \bibinfo {author} {\bibfnamefont {F.}~\bibnamefont
  {Verstraete}},\ }\bibfield  {title} {\bibinfo {title} {Matrix product states
  and projected entangled pair states: Concepts, symmetries, theorems},\ }\href
  {https://doi.org/10.1103/RevModPhys.93.045003} {\bibfield  {journal}
  {\bibinfo  {journal} {Rev. Mod. Phys.}\ }\textbf {\bibinfo {volume} {93}},\
  \bibinfo {pages} {045003} (\bibinfo {year} {2021})}\BibitemShut {NoStop}%
\bibitem [{\citenamefont {White}(1992)}]{White1992Density}%
  \BibitemOpen
  \bibfield  {author} {\bibinfo {author} {\bibfnamefont {S.~R.}\ \bibnamefont
  {White}},\ }\bibfield  {title} {\bibinfo {title} {Density matrix formulation
  for quantum renormalization groups},\ }\href
  {https://doi.org/10.1103/PhysRevLett.69.2863} {\bibfield  {journal} {\bibinfo
   {journal} {Phys. Rev. Lett.}\ }\textbf {\bibinfo {volume} {69}},\ \bibinfo
  {pages} {2863} (\bibinfo {year} {1992})}\BibitemShut {NoStop}%
\bibitem [{\citenamefont {White}(1993)}]{PhysRevB.48.10345}%
  \BibitemOpen
  \bibfield  {author} {\bibinfo {author} {\bibfnamefont {S.~R.}\ \bibnamefont
  {White}},\ }\bibfield  {title} {\bibinfo {title} {Density-matrix algorithms
  for quantum renormalization groups},\ }\href
  {https://doi.org/10.1103/PhysRevB.48.10345} {\bibfield  {journal} {\bibinfo
  {journal} {Phys. Rev. B}\ }\textbf {\bibinfo {volume} {48}},\ \bibinfo
  {pages} {10345} (\bibinfo {year} {1993})}\BibitemShut {NoStop}%
\bibitem [{\citenamefont {Johnson}\ \emph {et~al.}(2010)\citenamefont
  {Johnson}, \citenamefont {Clark},\ and\ \citenamefont
  {Jaksch}}]{Johnson2010Dynamical}%
  \BibitemOpen
  \bibfield  {author} {\bibinfo {author} {\bibfnamefont {T.~H.}\ \bibnamefont
  {Johnson}}, \bibinfo {author} {\bibfnamefont {S.~R.}\ \bibnamefont {Clark}},\
  and\ \bibinfo {author} {\bibfnamefont {D.}~\bibnamefont {Jaksch}},\
  }\bibfield  {title} {\bibinfo {title} {Dynamical simulations of classical
  stochastic systems using matrix product states},\ }\href
  {https://doi.org/10.1103/PhysRevE.82.036702} {\bibfield  {journal} {\bibinfo
  {journal} {Phys. Rev. E}\ }\textbf {\bibinfo {volume} {82}},\ \bibinfo
  {pages} {036702} (\bibinfo {year} {2010})}\BibitemShut {NoStop}%
\bibitem [{\citenamefont {Singh}\ \emph {et~al.}(2014)\citenamefont {Singh},
  \citenamefont {Pfeifer}, \citenamefont {Vidal},\ and\ \citenamefont
  {Brennen}}]{Singh2014Matrix}%
  \BibitemOpen
  \bibfield  {author} {\bibinfo {author} {\bibfnamefont {S.}~\bibnamefont
  {Singh}}, \bibinfo {author} {\bibfnamefont {R.~N.~C.}\ \bibnamefont
  {Pfeifer}}, \bibinfo {author} {\bibfnamefont {G.}~\bibnamefont {Vidal}},\
  and\ \bibinfo {author} {\bibfnamefont {G.~K.}\ \bibnamefont {Brennen}},\
  }\bibfield  {title} {\bibinfo {title} {Matrix product states for anyonic
  systems and efficient simulation of dynamics},\ }\href
  {https://doi.org/10.1103/PhysRevB.89.075112} {\bibfield  {journal} {\bibinfo
  {journal} {Phys. Rev. B}\ }\textbf {\bibinfo {volume} {89}},\ \bibinfo
  {pages} {075112} (\bibinfo {year} {2014})}\BibitemShut {NoStop}%
\bibitem [{\citenamefont {Mazza}\ \emph {et~al.}(2019)\citenamefont {Mazza},
  \citenamefont {Perfetto}, \citenamefont {Lerose}, \citenamefont {Collura},\
  and\ \citenamefont {Gambassi}}]{Mazza2019Suppression}%
  \BibitemOpen
  \bibfield  {author} {\bibinfo {author} {\bibfnamefont {P.~P.}\ \bibnamefont
  {Mazza}}, \bibinfo {author} {\bibfnamefont {G.}~\bibnamefont {Perfetto}},
  \bibinfo {author} {\bibfnamefont {A.}~\bibnamefont {Lerose}}, \bibinfo
  {author} {\bibfnamefont {M.}~\bibnamefont {Collura}},\ and\ \bibinfo {author}
  {\bibfnamefont {A.}~\bibnamefont {Gambassi}},\ }\bibfield  {title} {\bibinfo
  {title} {Suppression of transport in nondisordered quantum spin chains due to
  confined excitations},\ }\href {https://doi.org/10.1103/PhysRevB.99.180302}
  {\bibfield  {journal} {\bibinfo  {journal} {Phys. Rev. B}\ }\textbf {\bibinfo
  {volume} {99}},\ \bibinfo {pages} {180302(R)} (\bibinfo {year}
  {2019})}\BibitemShut {NoStop}%
\bibitem [{\citenamefont {Friedman}\ \emph {et~al.}(2022)\citenamefont
  {Friedman}, \citenamefont {Ware}, \citenamefont {Vasseur},\ and\
  \citenamefont {Potter}}]{Friedman2022Topologicaledge}%
  \BibitemOpen
  \bibfield  {author} {\bibinfo {author} {\bibfnamefont {A.~J.}\ \bibnamefont
  {Friedman}}, \bibinfo {author} {\bibfnamefont {B.}~\bibnamefont {Ware}},
  \bibinfo {author} {\bibfnamefont {R.}~\bibnamefont {Vasseur}},\ and\ \bibinfo
  {author} {\bibfnamefont {A.~C.}\ \bibnamefont {Potter}},\ }\bibfield  {title}
  {\bibinfo {title} {Topological edge modes without symmetry in
  quasiperiodically driven spin chains},\ }\href
  {https://doi.org/10.1103/PhysRevB.105.115117} {\bibfield  {journal} {\bibinfo
   {journal} {Phys. Rev. B}\ }\textbf {\bibinfo {volume} {105}},\ \bibinfo
  {pages} {115117} (\bibinfo {year} {2022})}\BibitemShut {NoStop}%
\bibitem [{\citenamefont {Urbanek}\ and\ \citenamefont
  {Sold{\'a}n}(2016)}]{URBANEK2016170}%
  \BibitemOpen
  \bibfield  {author} {\bibinfo {author} {\bibfnamefont {M.}~\bibnamefont
  {Urbanek}}\ and\ \bibinfo {author} {\bibfnamefont {P.}~\bibnamefont
  {Sold{\'a}n}},\ }\bibfield  {title} {\bibinfo {title} {Parallel
  implementation of the time-evolving block decimation algorithm for the
  bose-hubbard model},\ }\href
  {https://doi.org/https://doi.org/10.1016/j.cpc.2015.10.016} {\bibfield
  {journal} {\bibinfo  {journal} {Computer Physics Communications}\ }\textbf
  {\bibinfo {volume} {199}},\ \bibinfo {pages} {170} (\bibinfo {year}
  {2016})}\BibitemShut {NoStop}%
\bibitem [{\citenamefont {S\'anchez-Burillo}\ \emph {et~al.}(2017)\citenamefont
  {S\'anchez-Burillo}, \citenamefont {Zueco}, \citenamefont
  {Mart\'{\i}n-Moreno},\ and\ \citenamefont
  {Garc\'{\i}a-Ripoll}}]{PhysRevA.96.023831}%
  \BibitemOpen
  \bibfield  {author} {\bibinfo {author} {\bibfnamefont {E.}~\bibnamefont
  {S\'anchez-Burillo}}, \bibinfo {author} {\bibfnamefont {D.}~\bibnamefont
  {Zueco}}, \bibinfo {author} {\bibfnamefont {L.}~\bibnamefont
  {Mart\'{\i}n-Moreno}},\ and\ \bibinfo {author} {\bibfnamefont {J.~J.}\
  \bibnamefont {Garc\'{\i}a-Ripoll}},\ }\bibfield  {title} {\bibinfo {title}
  {Dynamical signatures of bound states in waveguide qed},\ }\href
  {https://doi.org/10.1103/PhysRevA.96.023831} {\bibfield  {journal} {\bibinfo
  {journal} {Phys. Rev. A}\ }\textbf {\bibinfo {volume} {96}},\ \bibinfo
  {pages} {023831} (\bibinfo {year} {2017})}\BibitemShut {NoStop}%
\bibitem [{\citenamefont {Qiao}\ \emph {et~al.}(2019)\citenamefont {Qiao},
  \citenamefont {Song},\ and\ \citenamefont {Sun}}]{Qiao2019Quantumphase}%
  \BibitemOpen
  \bibfield  {author} {\bibinfo {author} {\bibfnamefont {L.}~\bibnamefont
  {Qiao}}, \bibinfo {author} {\bibfnamefont {Y.-J.}\ \bibnamefont {Song}},\
  and\ \bibinfo {author} {\bibfnamefont {C.-P.}\ \bibnamefont {Sun}},\
  }\bibfield  {title} {\bibinfo {title} {Quantum phase transition and
  interference trapping of populations in a coupled-resonator waveguide},\
  }\href {https://doi.org/10.1103/PhysRevA.100.013825} {\bibfield  {journal}
  {\bibinfo  {journal} {Phys. Rev. A}\ }\textbf {\bibinfo {volume} {100}},\
  \bibinfo {pages} {013825} (\bibinfo {year} {2019})}\BibitemShut {NoStop}%
\bibitem [{SI()}]{SI}%
  \BibitemOpen
  \href@noop {} {\bibinfo  {journal} {Supplementary information for I.
  Constructing $|i\rangle_n$ in MPS form, II. Constructing $e^{-I{\hat H}\tau}$
  in MPO form, III. Time evolution, IV. Transmission and correlation function,
  and V. Time evolution of photon distribution}\ }\BibitemShut {NoStop}%
\bibitem [{\citenamefont {Suzuki}(1976)}]{suzuki1976generalized}%
  \BibitemOpen
\bibfield  {journal} {  }\bibfield  {author} {\bibinfo {author} {\bibfnamefont
  {M.}~\bibnamefont {Suzuki}},\ }\bibfield  {title} {\bibinfo {title}
  {Generalized trotter's formula and systematic approximants of exponential
  operators and inner derivations with applications to many-body problems},\
  }\href {https://link.springer.com/article/10.1007/BF01609348} {\bibfield
  {journal} {\bibinfo  {journal} {Commun. Math. Phys.}\ }\textbf {\bibinfo
  {volume} {51}},\ \bibinfo {pages} {183} (\bibinfo {year} {1976})}\BibitemShut
  {NoStop}%
\end{thebibliography}
%

\clearpage

\widetext

\section*{Supplementary information for ``Statistics of tens-of-photon states scattered by optical cavity, two-level atom and Jaynes-Cummings emitter"}

\section{Constructing $\left|i\right\rangle_n$ in MPS form}\label{sec1}

The $n$-photon initial state we aim to construct is denoted as Eq.~(6) in the main text, i.e., $\left|i\right\rangle_{n}= \sum_{l_1,l_2,\cdots, l_n}
\phi_{l_1}\phi_{l_2}\cdots \phi_{l_n}
\hat a_{l_1}^\dag \hat a_{l_2}^\dag \cdots \hat a_{l_n}^\dag
\left|\varnothing\right\rangle$.
Here, $\left|\varnothing\right\rangle$ represents the vacuum state with no excitation.
The coefficients $\phi_{l}$ follow a Gaussian profile,
$
\phi_l=N\sum_{k}e^{-(k-k_{0})^{2}/k_{w}^{2}}e^{-ik(x_l-x_0)},
$
where the wave vector $k={2\pi\over L}l$ ($l=1, 2, \cdots, L$). The Hilbert space of the system states grows exponentially as the photon number increases. In this section, we introduce the specific procedures for constructing the $n$-photon initial state $\left|i\right\rangle_n$ in the MPS (Matrix Product State) form.

\subsection{Constructing straightly}

Let's consider a Fock state with $n$ photons distributed among $m$ sites. These sites are labeled by $i_1$, $i_2$, $\cdots$, and $i_m$, with the corresponding occupation states denoted as $|n_{i_1},n_{i_2},\cdots,n_{i_m}\rangle$. The MPS representation of $|n_{i_1},n_{i_2},\cdots,n_{i_m}\rangle$ reads
\begin{align}
|n_{i_1},n_{i_2},\cdots,n_{i_m}\rangle
=\sum_{\{\sigma\}}M_{1}^{[n]\sigma_1}M_{2}^{[n]\sigma_2}\cdots M_{L}^{[n]\sigma_L} |\sigma_1,\sigma_2\cdots\sigma_L\rangle,
\end{align}
where the total photon number $n=\sum_{\alpha=1}^{m} n_{i_\alpha}$.
The matrices have the form
\begin{align}
\begin{array}{ll}
M_{l}^{[n]\sigma_l}=\left\{\begin{array}{cl}
                         1,&~~\sigma_l=n_l,\\
                         0,&~~\sigma_l\neq n_l,
                       \end{array}\right.
&{\rm for~} l\in\{i_1,i_2,\cdots,i_m\},\\
M_{l}^{[n]\sigma_l}=\left\{\begin{array}{cl}
                         1,&~~\sigma_l=0,\\
                         0,&~~\sigma_l\neq 0,
                       \end{array}\right.
& {\rm for~} l\notin\{i_1,i_2,\cdots,i_m\}.
\end{array}
\end{align}

Considering an example that there are $3$ photons in the cavity chain, it has three types of distributions: all $3$ photons in one site, $2$ photons in one site and $1$ photon in another one, and $3$ photons in three different sites.
For the first type, the site matrices for the state $|3_{i}\rangle$ can be written as
\begin{align}
\begin{array}{rrrr}
  M_i^{[3]0}=0, & M_i^{[3]1}=0, & M_i^{[3]2}=0, & M_i^{[3]3}=1, \\
  M_{l\neq i}^{[3]0}=1, & M_{l\neq i}^{[3]1}=0, & M_{l\neq i}^{[3]2}=0, & M_{j\neq i}^{[3]3}=0.
\end{array}
\end{align}
For the second type $|2_{i_1},1_{i_2}\rangle$, the photons occupy the sites $i_1$ and $i_2$. The matrices for $|2_{i_1},1_{i_2}\rangle$ are
\begin{align}
\begin{array}{rrrr}
  M_{i_1}^{[3]0}=0, & M_{i_1}^{[3]1}=0, & M_{i_1}^{[3]2}=1, & M_{i_1}^{[3]3}=0, \\
  M_{i_2}^{[3]0}=0, & M_{i_2}^{[3]1}=1, & M_{i_2}^{[3]2}=0, & M_{i_2}^{[3]3}=0, \\
  M_{l\neq i_1,i_2}^{[3]0}=1, & M_{l\neq i_1,i_2}^{[3]1}=0, & M_{l\neq i_1,i_2}^{[3]2}=0, & M_{l\neq i_1,i_2}^{[3]3}=0.
\end{array}
\end{align}
Similarly, the site matrices for the third type $|1_{i_1},1_{i_2},1_{i_3}\rangle$ are
\begin{align}
\begin{array}{rrrr}
  M_{i_1}^{[3]0}=0, & M_{i_1}^{[3]1}=1, & M_{i_1}^{[3]2}=0, & M_{i_1}^{[3]3}=0, \\
  M_{i_2}^{[3]0}=0, & M_{i_2}^{[3]1}=1, & M_{i_2}^{[3]2}=0, & M_{i_2}^{[3]3}=0, \\
  M_{i_3}^{[3]0}=0, & M_{i_3}^{[3]1}=1, & M_{i_3}^{[3]2}=0, & M_{i_3}^{[3]3}=0, \\
  M_{l\neq i_1,i_2,i_3}^{[3]0}=1, & M_{l\neq i_1,i_2,i_3}^{[3]1}=0, & M_{l\neq i_1,i_2,i_3}^{[3]2}=0, & M_{l\neq i_1,i_2,i_3}^{[3]3}=0.
\end{array}
\end{align}

Using the above method, we are able to construct the MPS for arbitrary Fock states and thus their superposition.
Let's take the superposition of $\cal J$ MPSs, named $|a_j\rangle$, as
\begin{align}
|A\rangle&=\sum_{j=1}^{\cal{J}}C_j|a_j\rangle,
 \end{align}
where the MPS for the Fock state $|a_j\rangle$ has the form
 \begin{align}
|a_j\rangle=\sum_{\{\bm \sigma\}}M_{[j]1}^{[n]\sigma_1}M_{[j]2}^{[n]\sigma_2}\cdots
M_{[j]L}^{[n]\sigma_L}|\sigma_1,\sigma_2,\cdots,\sigma_L\rangle.
\end{align}
Here the subscript $[j]$ denotes the $j$th Fock state.
The MPS for the superposition state, denoted as
\begin{align}
|A\rangle
=\sum_{\{\bm \sigma\}}A_{1}^{[n]\sigma_1}A_{2}^{[n]\sigma_2}\cdots
A_{L}^{[n]\sigma_L}|\sigma_1,\sigma_2,\cdots,\sigma_L\rangle,
 \end{align}
can be obtained by the following transformation
\begin{align}
A_{l}^{[n]\sigma_l}=\left\{\begin{array}{ll}
\qquad \left(C_1^{\frac{1}{L}} M_{[1]l}^{[n]\sigma_l},C_2^{1\over L} M_{[2]l}^{[n]\sigma_l},\cdots,C_{\cal J}^{\frac{1}{L}} M_{[{\cal J}]l}^{[n]\sigma_l}\right),& l=1,\\
{\rm diag}\left(C_1^{\frac{1}{L}} M_{[1]l}^{[n]\sigma_l},C_2^{1\over L} M_{[2]l}^{[n]\sigma_l},\cdots,C_{\cal J}^{\frac{1}{L}} M_{[{\cal J}]l}^{[n]\sigma_l}\right),& l\neq1~ {\rm or}~L,\\
\qquad \left(C_1^{\frac{1}{L}} M_{[1]l}^{[n]\sigma_l},C_2^{1\over L} M_{[2]l}^{[n]\sigma_l},\cdots,C_{\cal J}^{\frac{1}{L}} M_{[{\cal J}]l}^{[n]\sigma_l}\right)^T,& l=L.
\end{array}\right.
\end{align}
Hence, the MPS for $\left|i\right\rangle_{n}= \sum_{l_1,l_2,\cdots, l_n}
\phi_{l_1}\phi_{l_2}\cdots \phi_{l_n}
\hat a_{l_1}^\dag \hat a_{l_2}^\dag \cdots \hat a_{l_n}^\dag
\left|\varnothing\right\rangle$ can be constructed in principle.
However, the aforementioned approach becomes impractical in real operations due to the often large number of Fock states involved in a superposition.
For example, the Gaussian state we considered in the main text.
If there are $3$ photons in the waveguide with $L=256$, the number of Fock states is $C_L^1+2C_L^2+C_L^3=2\ 796\ 416$.
As a result, it becomes essential to find an efficient method for constructing the MPS, i.e., the iterative approach described in the main text.

\subsection{Iteration method for constructing MPS}\label{sec1B}

In the case of the Gaussian state discussed in the main text, as given in Eq.~(6), the iterative method relies on the relationship $\left|i\right\rangle_1=\sum_{l_1}\phi_{l_1}\hat{a}^{\dag}_{l_1}\left|\varnothing\right\rangle$.
The detailed procedures are presented following.

Let's start by constructing the MPS for $\left|i\right\rangle_1$. 
At first, we write the single-occupation state $\phi_{l_1}\hat a_{l_1}^\dag|\varnothing\rangle$ into the MPS by taking its matrix set $\left\{M_{[l_1]l}^{[1]\sigma_l}\right\}$ as
\begin{align}\label{Mn1}
M_{[l_1]l}^{[1]0}=\left\{\begin{array}{cc}
1, & l\neq l_1 \\
0, & l = l_1
\end{array}\right.\!\!\!,\ \ \ \
M_{[l_1]l}^{[1]1}=\left\{\begin{array}{cc}
0, & l\neq l_1 \\
\phi_{l_1}, & l = l_1
\end{array}\right.\!\!\!.
\end{align}
For the superposition state $\left| i\right\rangle_1=\sum_{l_1}\phi_{l_1}\hat{a}^{\dag}{l_1}\left|\varnothing\right\rangle$, the matrix set $\left\{M_{l}^{[1]\sigma_l}\right\}$ of the MPS representation can be written as:
\begin{align}\label{superposition state 1}
M_{l}^{[1]\sigma_l}=\left\{\begin{array}{ll}
\qquad\left(M_{[1]l}^{[1]\sigma_l}, M_{[2]l}^{[1]\sigma_l}, \cdots, M_{[L]l}^{[1]\sigma_l}\right), & l=1,\\
{\rm diag}\left(M_{[1]l}^{[1]\sigma_l}, M_{[2]l}^{[1]\sigma_l}, \cdots, M_{[L]l}^{[1]\sigma_l}\right),& l\neq1~{\rm or}~L,\\
\qquad\left(M_{[1]l}^{[1]\sigma_l}, M_{[2]l}^{[1]\sigma_l}, \cdots, M_{[L]l}^{[1]\sigma_l}\right)^T,& l=L,
\end{array}\right.
\end{align}
That is,
\begin{align}\label{single-occupation state}
|i\rangle_1=\sum_{\{\sigma\}}M_{1}^{[1]\sigma_1}M_{2}^{[1]\sigma_2}\cdots M_{L}^{[1]\sigma_L}|\sigma_1,\sigma_2,\cdots,\sigma_L\rangle.
\end{align}
Note that the superposition coefficients have been involved in
$\left\{M_{[l_1],l}^{[1]\sigma_l}\right\}$, see Eq.~(\ref{Mn1}).

The information provided thus far pertains to single-occupation states. For multi-occupation states, one can derive the matrix set $\left\{M_{[l_n]l}^{[n]\sigma_l}\right\}$ for the state $\phi_{l_n}\hat{a}^{\dag}_{l_n}\left|i\right\rangle_{n-1}$ from the matrix set $\left\{M_{l}^{[n-1]\sigma_l}\right\}$ associated with the state $\left|i\right\rangle_{n-1}$, using the relations
\begin{align}
&M_{[l_n]l}^{[n]\sigma_l}{\xlongequal{l=l_n}}\left\{\begin{array}{cc}
\phi_{l_n}\sqrt{\sigma_l}M_{l}^{[n-1]\sigma_l-1}, & 1\leq\sigma_l \leq n,\\
0, & \sigma_l = 0,
\end{array}\right.\\
&M_{[l_n]l}^{[n]\sigma_l}{\xlongequal{l\neq l_n}}\left\{\begin{array}{cc}
0, & \qquad\ \sigma_l = n,\\
M_{l}^{[n-1]\sigma_l}, & \qquad\ 0\leq\sigma_l \leq n-1.
\end{array}\right.
\end{align}
Subsequently, the MPS representation for the superposition state involving $\phi_{l_n}\hat{a}^{\dag}_{l_n}\left|i\right\rangle_{n-1}$ can be obtained through a procedure similar to what was done in Eq.~\eqref{superposition state 1}. In other words,
\begin{align}\label{superposition state n}
M_{l}^{[n]\sigma_l}=\left\{\begin{array}{ll}
\qquad(M_{[1]l}^{[n]\sigma_l}, M_{[2]l}^{[n]\sigma_l}, \cdots, M_{[L]l}^{[n]\sigma_l}), & l=1,\\
{\rm diag}(M_{[1]l}^{[n]\sigma_l}, M_{[2]l}^{[n]\sigma_l}, \cdots, M_{[L]l}^{[n]\sigma_l}),& l\neq1~{\rm or}~L,\\
\qquad(M_{[1]l}^{[n]\sigma_l}, M_{[2]l}^{[n]\sigma_l}, \cdots, M_{[L]l}^{[n]\sigma_l})^T,& l=L.
\end{array}\right.
\end{align}
Given that we have already established the MPS representation for the single-occupation state $\left|i\right\rangle_1$ as shown in Eq.~\eqref{single-occupation state}, we can proceed to derive the $n$-photon state $\left|i\right\rangle_n$ incrementally, step by step.

Using the aforementioned method, we can construct the MPS for the $n$-photon initial state $\left|i\right\rangle_n$ for the cavity chain, excluding the quantum emitter. 
Now, let's discuss how to incorporate the physical dimension of the quantum emitter.
For each site $S$, we treat the $S$th cavity and the quantum emitter as a unified system. 
In the initial state, we assume that the quantum emitter is neither excited nor occupied. 
Consequently, the state space expands from ${\sigma_{S}}$ to ${(\sigma_{S},\sigma_{c})}$, ${(\sigma_{S},\sigma_{a})}$ or ${(\sigma_{S},\sigma_{c},\sigma_{a})}$ when considering an OC, a TLA or a JCE, respectively .
Here, $\sigma_{S}$ represents the photon occupation number in the $S$th cavity of the chain, $\sigma_{c}$ denotes the photon occupation number in the OC, and $\sigma_{a}$ indicates whether the TLA is excited or not.
For the OC, when $\sigma_{c}=0$, we have $M_{S}^{[n](\sigma_{S},\sigma_{c})}=M_{S}^{[n]\sigma_S}$, otherwise $M_{S}^{[n](\sigma_{S},\sigma_{c})}=0$.
And for the TLA, when $\sigma_{a}=0$, we have $M_{S}^{[n](\sigma_{S},\sigma_{a})}=M_{S}^{[n]\sigma_S}$, otherwise $M_{S}^{[n](\sigma_{S},\sigma_{a})}=0$.
What's more, for the JCE, when $\sigma_{c}=\sigma_{a}=0$, we have $M_{S}^{[n] (\sigma_S,\sigma_c,\sigma_a)} =M_{S}^{[n]\sigma_S}$, otherwise, $M_{S}^{[n](\sigma_S,\sigma_c,\sigma_a)}=0$.
Through this approach, we can obtain the MPS representation for the $n$-photon initial state $\left|i\right\rangle_n$ as illustrated in Fig.~1(b).

It's crucial to recognize that the dimensions of the matrices will grow exponentially as we perform the procedure $\left|i\right\rangle_{n-1}\rightarrow\left|i\right\rangle_{n}$. 
Therefore, it becomes necessary to compress the matrices, typically achieved through Singular Value Decomposition (SVD) at each step.
Moreover, for computational efficiency, it's highly advantageous to transform the MPS into either the left-canonical or right-canonical form during the process. 
This canonicalization simplifies various calculations and reduces the computational cost associated with MPS-based methods.

\section{Constructing $e^{-i\hat{H}\tau}$ in MPO form}

In this section, we will outline the process of constructing the MPO for $e^{-i\hat{H}\tau}$ as given in Eq.~(7).

We begin with the second-order Trotter decomposition:
\begin{align}\label{Trotter2}
e^{-i\hat H\tau} =
e^{-i\hat H_{\rm odd}\tau/2}
e^{-i\hat H_{\rm even}\tau}
e^{-i\hat H_{\rm odd}\tau/2}+O(\tau^3),
\end{align}
where $\hat H_{\rm odd}$ and $\hat H_{\rm even}$ represent the Hamiltonians acting on the odd and even bonds, i.e., 
\begin{align}\label{MPO}
\hat H_{\rm odd} = \sum_{l\in \rm odd} \hat h_l, \qquad
\hat H_{\rm even} = \sum_{l\in \rm even} \hat h_l,
\end{align}
respectively, where
\begin{align}\label{hhl}
\hat{h}_{l}=&\frac{1}{2}\omega_{W}\!\!\left(
\hat{a}_{l}^{\dag}\hat{a}_{l}+\hat{a}_{l+1}^{\dag}\hat{a}_{l+1}
\right)
-\frac{1}{2}J\!\!\left(\hat{a}_{l}^{\dag}\hat{a}_{l+1}+\rm h.c.\right)
\nonumber\\
+&\!\left\{\begin{array}{lc}
\!\!\!\frac{1}{2}\omega_{W}\!\!\left(
\hat{a}_{1}^{\dag}\hat{a}_{1}\delta_{l,1}
+\hat{a}_{L}^{\dag}\hat{a}_{L}\delta_{l,L-1}\right), &l\neq S{-}1, S,\\
\!\!\!\frac{1}{2}\left(\hat{H}_{E}+\hat{H}_{I}\right), &l= S{-}1, S.
\end{array}\right.
\end{align}Because of $[\hat{h}_{l},\hat{h}_{l+2}]=0$, the equations
$
e^{-i\hat{H}_{\rm odd}\tau/2}=e^{-i\hat{h}_{1}\tau/2}e^{-i\hat{h}_{3}\tau/2}\cdots
$
and
$
e^{-i\hat{H}_{\rm even}\tau}=e^{-i\hat{h}_{2}\tau}e^{-i\hat{h}_{4}\tau}\cdots
$
are strict, whose last terms depend on the parity of $L$.
After substituting the bond Hamiltonian $\hat{h}_l$ (as seen in Eq.~\ref{hhl}) into $\hat H_{\rm odd}$ and $\hat H_{\rm even}$, our objective is to find the MPO for operators of the form $e^{-i\hat{h}_{l}\tau}$, which can be solved through the following steps.
Firstly, calculating the matrix representation of $e^{-i\hat{h}_{l}\tau}$ on the basis set of $\{\left|\sigma_{l},\sigma_{l+1}\right\rangle\}$, denoted as $D^{[n]}_{(\sigma_{l}\sigma_{l+1}),(\sigma'_{l}\sigma'_{l+1})}=\left\langle\sigma_l,\sigma_{l+1}\right|e^{-i\hat{h}_{l}\tau}\left|\sigma'_l,\sigma'_{l+1}\right\rangle$, where $\sigma_l$ is the physical index of the $l$th site.
Then, exchanging its second and third subscripts to get $\bar D^{[n]}_{(\sigma_{l}\sigma'_{l}),(\sigma_{l+1}\sigma'_{l+1})}$.
Thirdly, decomposing $\bar D^{[n]}_{(\sigma_{l}\sigma'_{l}),(\sigma_{l+1}\sigma'_{l+1})}$ into the form
\begin{align}
\bar D^{[n]}_{(\sigma_{l}\sigma'_{l}),(\sigma_{l+1}\sigma'_{l+1})}
= \sum_{d_s=1}^{D_s}U^{[n]}_{(\sigma_{l}\sigma'_{l}),d_s}
V^{[n]\dag}_{d_s,(\sigma_{l+1}\sigma'_{l+1})},
\end{align}
by the SVD, where $D_s$ is the number of the nonzero singular values.
Note that the singular values have been moved into $U$ and/or $V$.
Finally, the MPOs on the $l$th and $(l+1)$th sites can be obtained by reshaping $ U^{[n]}$ and $ V^{[n]\dag}$, that is,
$${U}^{[n]} \rightarrow
\left(\!Q^{[n]\sigma_{l}\sigma'_{l}}_l\!\right)_{1,D_s}\!\!, \quad
{V}^{[n]\dag} \rightarrow
\left(\!Q^{[n]\sigma_{l+1}\sigma'_{l+1}}_{l+1}\!\right)_{D_s\!, 1}\!\!.
$$
According to that the MPO for $e^{-i\hat{h}_{l}\tau'}$ can be expressed as:
\begin{align}\label{hl}
e^{-i\hat{h}_{l}\tau'}=\sum_{\{\sigma_l\sigma_{l+1}\}}
\sum_{\{\sigma'_{l}\sigma'_{l+1}\}}
Q_{l}^{[n]\sigma_l\sigma'_{l}}
Q_{l+1}^{[n]\sigma_{l+1}\sigma'_{l+1}}
\left|\sigma_l,\sigma_{l+1}\right\rangle
\left\langle\sigma'_l,\sigma'_{l+1}\right|.
\end{align}
Here, we distinguish two cases for $\tau'$: it is either $\tau/2$ or $\tau$ depending on whether $l$ belongs to the set of even or odd indices.
Using the expression in Eq.~\eqref{hl}, we can directly derive the MPO for both $e^{-i\hat{H}_{{\rm odd}}\tau/2}$ and $e^{-i\hat{H}_{{\rm even}}\tau}$,
that is,
\begin{align}
e^{-i\hat{H}_{{\rm odd}}\tau/2}&=\sum_{\{\bm \sigma, \bm \sigma'\}}
Q_{{\rm odd},1}^{[n]\sigma_{1}\sigma'_{1}}Q_{{\rm odd},2}^{[n]\sigma_{2}\sigma'_{2}}
\cdots
Q_{{\rm odd},L}^{[n]\sigma_{L}\sigma'_{L}}
\left|\sigma_{1}\sigma_{2}\cdots\sigma_{L}\right\rangle
\left\langle\sigma'_{1}\sigma'_{2}\cdots\sigma'_{L}\right|,\\
e^{-i\hat{H}_{{\rm odd}}\tau}&=\sum_{\{\bm \sigma, \bm \sigma'\}}
Q_{{\rm even},1}^{[n]\sigma_{1}\sigma'_{1}}Q_{{\rm even},2}^{[n]\sigma_{2}\sigma'_{2}}
\cdots
Q_{{\rm even},L}^{[n]\sigma_{L}\sigma'_{L}}
\left|\sigma_{1}\sigma_{2}\cdots\sigma_{L}\right\rangle
\left\langle\sigma'_{1}\sigma'_{2}\cdots\sigma'_{L}\right|.
\end{align}
Then the matrices of MPO of Eq.~\eqref{Trotter2} can be written as
\begin{align}
Q_{l}^{[n]\sigma_l\sigma'''_{l}}=
\sum_{\sigma'_l\sigma''_l}
Q_{{\rm odd},l}^{[n]\sigma_l\sigma'_{l}}\otimes
Q_{{\rm even},l}^{[n]\sigma'_l\sigma''_{l}}\otimes
Q_{{\rm odd},l}^{[n]\sigma''_l\sigma'''_{l}},
\end{align}
where $\otimes$ represents the Kronecker product between matrices.

In this way, the MPO of $e^{-i\hat{H}\tau}$ is successfully constructed, namely,
\begin{align}
e^{-i\hat{H}\tau}=\sum_{\{\bm \sigma, \bm \sigma'\}}
Q_{1}^{[n]\sigma_{1}\sigma'_{1}}Q_{2}^{[n]\sigma_{2}\sigma'_{2}}
\cdots
Q_{L}^{[n]\sigma_{L}\sigma'_{L}}
\left|\sigma_{1}\sigma_{2}\cdots\sigma_{L}\right\rangle
\left\langle\sigma'_{1}\sigma'_{2}\cdots\sigma'_{L}\right|.
\end{align}

\section{Time evolution}

Since the $n$-photon initial MPS $|i\rangle_n$ and time evolution MPO $e^{-i\hat{H}\tau}$ are constructed, the MPS of $|\Phi(t+\tau)\rangle$ at time $t+\tau$ can be obtained by applying the MPO to the MPS of $|\Phi(t)\rangle$ at time $t$.
The method of direct application \cite{PAECKEL2019167998} is used in this work, i.e.,
\begin{align}
\hat{O}_\tau^{[n]}|\Phi(t)\rangle_n&=
\sum_{\{\bm{\sigma,\sigma'}\}}\sum_{\{{\bm \alpha}\}}Q_{1;\alpha_0,\alpha_1}^{[n]\sigma_1\sigma'_1} Q_{2;\alpha_1,\alpha_2}^{[n]\sigma_2\sigma'_2}\cdots Q_{L;\alpha_{L-1},\alpha_L}^{[n]\sigma_L\sigma'_L}|\sigma_1,\sigma_2,\cdots, \sigma_L\rangle\langle\sigma'_1,\sigma'_2,
\cdots,\sigma'_L|\nonumber\\
&~~~~\times \sum_{\{\bm\sigma''\}}\sum_{\{{\bm \beta}\}} M_{1;\beta_0,\beta_1}^{[n],\sigma''_1}
M_{2;\beta_1,\beta_2}^{[n],\sigma''_2}\cdots M_{L;\beta_{L-1},\beta_L}^{[n],\sigma''_L}|\sigma''_1,\sigma''_2,\cdots,\sigma''_L\rangle\nonumber\\
&=\sum_{\{\bm\sigma,{\bm \gamma}\}}
M_{1;\alpha_0\beta_0,\alpha_1\beta_1}^{\prime[n]\sigma_1}
M_{2;\alpha_1\beta_1,\alpha_2\beta_2}^{\prime[n]\sigma_2}\cdots M_{L;\alpha_{L-1}\beta_{L-1},\alpha_{L}\beta_{L}}^{\prime[n]\sigma_L}|\sigma_1,\sigma_2,\cdots,\sigma_L\rangle\nonumber\\
&=|\Phi(t+\tau)\rangle_n.
\end{align}
The matrix $M^{\prime[n]\sigma_l}_l$ is given by
\begin{align}
M_{l;\alpha_{l-1}\beta_{l-1},\alpha_{l}\beta_{l}}^{\prime[n]\sigma_l}
=\sum_{\sigma'_l}Q_{l;\alpha_{l-1},\alpha_l}^{[n]\sigma_l\sigma'_l}
M_{l;\beta_{l-1},\beta_l}^{[n],\sigma'_l}.
\end{align}
As we can see, the bond dimension of the matrices in the MPS increases every time when applying the MPO to it. 
Therefore, it's crucial to perform bond dimension compression using techniques like SVD after each step of time evolution to manage the increasing computational complexity.

\section{Transmission and correlation function}

By combining the information provided in the main text and the preceding sections, the final state $\left|f\right\rangle_n$ can be accurately calculated through time evolution. The photon number on the $l$th site can be determined as follows,
\begin{align}\label{photon distribution}
n_l={}_{n}\! \langle f|\hat{n}_l|f\rangle_n,
\end{align}
where $\hat{n}_l=\hat{a}_{l}^{\dag}\hat{a}_l$ is the particle number operator.
To simplify calculation, we transform the final state into a mixed-canonical MPS \cite{PAECKEL2019167998}, that is,
\begin{align}\label{mps}
\left|f\right\rangle_{n}=
\sum_{\{\bm \sigma\}}A^{[n]\sigma_{1}}_{1}A^{[n]\sigma_{2}}_{2}\cdots
A^{[n]\sigma_{l-1}}_{l-1} M^{[n]\sigma_{l}}_{l}
B^{[n]\sigma_{l+1}}_{l+1}\cdots
B^{[n]\sigma_{L-1}}_{L-1}B^{[n]\sigma_{L}}_{L} \left|\sigma_{1}\sigma_{2}\cdots\sigma_{L}\right\rangle.
\end{align}
Except the matrix $M$, all $A$ ($B$) are left-normalized (right-normalized), i.e.,
\begin{align} \label{noemalized-tensor}
\sum_{\sigma_l}A_{l}^{[n]\sigma_l\dag} A_{l}^{[n]\sigma_l}=I
,\qquad
\sum_{\sigma_l}B_{l}^{[n]\sigma_l}B_{l}^{[n]\sigma_l\dag}=I,
\end{align}
where $I$ is a unitary matrix.
Then using Eq.~(\ref{noemalized-tensor}), one can get
\begin{align}
n_l={\rm Tr}\left(\sum_{\sigma_l}
M^{[n]\sigma_l\dag}_{l}\sigma_l
M^{[n]\sigma_l}_{l}\right).
\end{align}
Accordingly, we can arrive at the transmission $T$ by
\begin{align}
T=\frac{\sum_{l=S+1}^{L}n_l}{\sum_{l=1}^{L}n_l}.
\end{align}

In the main text, we use the JCE as an example to compute the transmission and present the results in Fig.~2. When the Rabi coupling $\Omega$ is set to 0, the quantum emitter effectively becomes an OC.

Next, we will describe how to compute the second-order correlation function $G^{(2)}_{l,l'}={}_{n}\! \left\langle f\right|\hat{a}^{\dag}_{l}\hat{a}^{\dag}_{l'}\hat{a}_{l'} \hat{a}_{l}\left|f\right\rangle_{n}$, which obeys the symmetry property $G^{(2)}_{l,l'}=G^{(2)}_{l',l}$.
For convenience in calculation, we will transform the final state into the following mixed-canonical form,
\begin{align}
\left|f\right\rangle_{n}=
\sum_{\{\bm \sigma\}}A^{[n]\sigma_{1}}_{1}\cdots A^{[n]\sigma_{l-1}}_{l-1} M^{[n]\sigma_{l}}_{l}M^{[n]\sigma_{l+1}}_{l+1}\cdots M^{[n]\sigma_{l'}}_{l'} B^{[n]\sigma_{l'+1}}_{l'+1}\cdots B^{[n]\sigma_{L}}_{L}\left|\sigma_{1}\sigma_{2}\cdots\sigma_{L}\right\rangle,
\end{align}
where we assume $l'\geq l$.
Similar to Eq.~\eqref{mps}, all $A$ ($B$) are left-normalized (right-normalized), except the matrices $M$.
Then we arrive at
\begin{align}
G^{(2)}_{l,l'}=\left\{\begin{array}{ll}
{\rm Tr}\sum_{\{\sigma_l,\cdots \sigma_l'\}}\left(
M_{l'}^{[n]\sigma_{l'}\dag}
\cdots M_{l}^{[n]\sigma_{l}\dag}\sigma_l\sigma_{l'}
M_{l}^{[n]\sigma_{l}}\cdots M_{l'}^{[n]\sigma_{l'}}\right),&
l\neq l',\\
{\rm Tr}\sum_{\{\sigma_l\}}\left(
M_{l}^{[n]\sigma_{l}\dag}\sigma_l(\sigma_l-1)
M_{l}^{[n]\sigma_{l}}\right),& l=l'.
\end{array}\right.
\end{align}
Consequently, the normalized second order correlation function $g_{l}^{(2)}$ can be obtained by
\begin{align}
g_{l}^{(2)}=\frac{G^{(2)}_{l,l}}{n_l^2}.
\end{align}
\begin{figure}[h]\label{FigS1}
  \centering
  \includegraphics[width=0.8\textwidth]{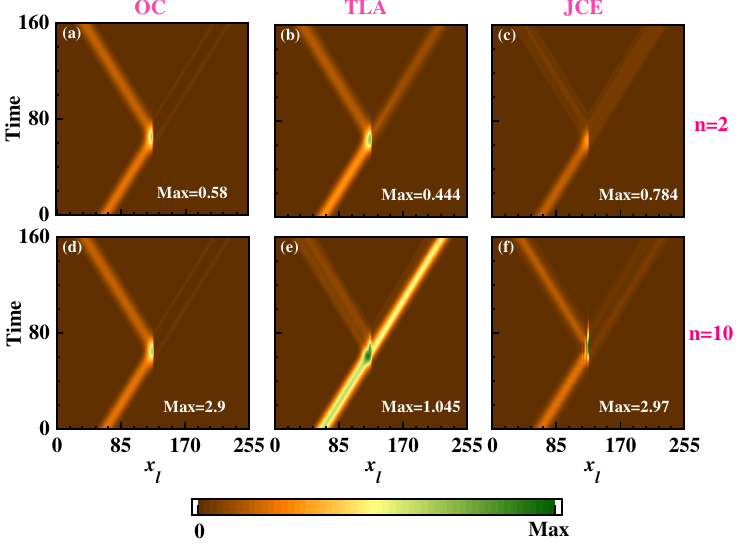}\\
  \caption{Variations of the photon number distributions with time that scattered by OC (a, d), TLA (b, e) and JCE (c, f).
  Where the total photon number $n=2$ ($n=10$) for the first (second) row.
  Parameters: $L=256$, $S=128$, $k_{w}=0.05\pi/d$, $\varepsilon_0=0$, $x_{0}=64d$, $\omega_{a}=\omega_{c}=\omega_{W}\equiv0$, $V_0=0.4J$, and $\Omega=0.15J$.}\label{FigS1}
\end{figure} 

\section{Time evolution of photon distribution}

In the main text, the Hamiltonian for the system with a JCE has been given, and when the Rabbi coupling
 $\Omega=0$ we can achieve that of OC.
Further more, for the case of TLA, the Hamiltonian is described as follows,
\begin{align}
\hat{H}_{W}&=\sum_{l=1}^{L}\omega_{W}\hat{a}_{l}^{\dag}\hat{a}_{l}-\frac{J}{2} \sum_{l=1}^{L-1}\left(\hat{a}_{l}^{\dag}\hat{a}_{l+1}+\rm h.c.\right),\\
\hat{H}_E&=\omega_a\hat{\sigma}^{+}\hat{\sigma}^{-},\\
\hat{H}_{I}&=V_0\left(\hat{a}_{s}^{\dag}\hat{\sigma}^{-} +\hat{a}_{s}\hat{\sigma}^{+}\right).
\end{align}

According to the previous descriptions, we calculate and plot the variations of $n_l$ in each site with time for the cases of $n=2$ and $n=10$ in Fig.~\ref{FigS1}.
From Fig.~\ref{FigS1} (a, d), we can clearly see that the photons' transmission scattered by the OC is independent on $n$.
However, unlike the OC, the TLA can only absorb one photon one time, which leads to the dependence of the transmission on photon number $n$, as showed in Fig.~\ref{FigS1} (b, e).
The transmissions increase as $n$ increases.
On the contrary, the transmissions decrease as $n$ increases in Fig.~\ref{FigS1} (c, f).
The reason is that the larger the $n$ is, the JCE acts more like an OC.
The detailed analyses for the photon statistics of the final states are in the main text.

\end{document}